\documentclass[12pt]{article}

\usepackage[
    a4paper,
    margin=1in,
    headheight=64pt,
    headsep=16pt
]{geometry}

\usepackage[T1]{fontenc}
\usepackage[utf8]{inputenc}
\usepackage{newtxtext,newtxmath}

\usepackage{graphicx}
\usepackage{booktabs}
\usepackage{tabularx}
\usepackage{array}
\usepackage{multirow}
\usepackage{amsmath}
\usepackage{siunitx}
\usepackage{xcolor}
\usepackage{setspace}
\usepackage{microtype}
\usepackage{caption}
\usepackage{float}
\usepackage{pdflscape}
\usepackage{placeins}
\usepackage{lineno}
\usepackage{enumitem}
\usepackage{longtable}
\usepackage{threeparttable}
\usepackage{makecell}
\usepackage{eso-pic}

\usepackage[
    backend=biber,
    style=numeric-comp,
    sorting=none
]{biblatex}

\usepackage[hidelinks]{hyperref}

\usepackage{fancyhdr}

\fancypagestyle{plain}{
    \fancyhf{}

    \fancyhead[L]{%
        \raisebox{-0.10cm}{%
            \includegraphics[
                width=0.28\textwidth
            ]{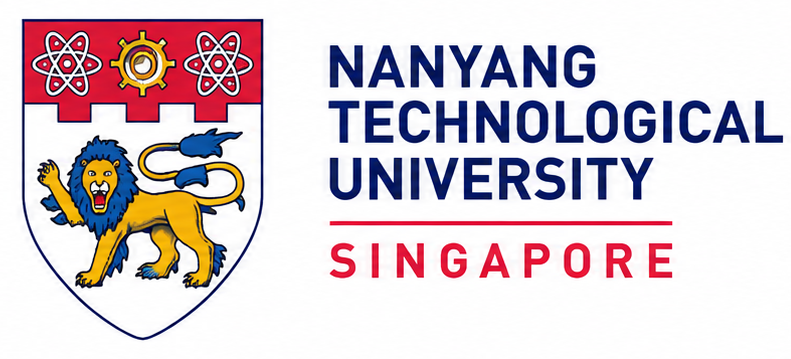}%
        }%
    }

    \fancyhead[R]{%
        \raisebox{-0.10cm}{%
            \includegraphics[
                width=0.38\textwidth
            ]{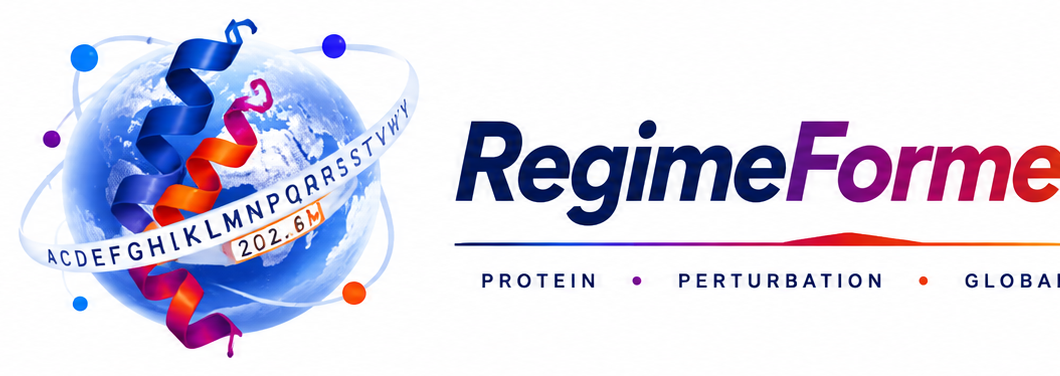}%
        }%
    }

    \fancyfoot[C]{\thepage}
}

\newcommand{\RegimeFormer}{RegimeFormer}
\newcommand{\RegimeAtlas}{RegimeAtlas}

\definecolor{TitlePanel}{RGB}{241,242,250}
\definecolor{TitleInk}{RGB}{28,44,43}

\newcommand{\PaperTitle}{
RegimeFormer: A Large Protein Model of Global Perturbation Regimes
}

\newcommand{\PaperAuthors}{%
Siyuan Ma$^{1,\dagger}$,
Yi Chai$^{2,\dagger}$,
Yi Wu$^{3,\dagger}$,
Qixin Zhang$^{1,\dagger}$,
Yajing Yuan$^{4,\dagger}$,
Kanglu Zhao$^{5,\dagger}$,
Zhikang Chen$^{6,\dagger}$,
Haowei Wang$^{7,\dagger}$,
Shuying Cao$^{8,\dagger}$,
Xiaolei Yu$^{9,*}$,
Xiangfei Han$^{10,\dagger}$,
Yun Liu$^{11,\dagger}$,
Yang Liu$^{1,\dagger}$,
Tingting Zhu$^{6,\dagger}$,
Dacheng Tao$^{1,*}$
}

\begin{document}

\begin{titlepage}

\thispagestyle{fancy}

%
\AddToShipoutPictureBG*{%
    \AtTextLowerLeft{%
        \color{TitlePanel}%
        \rule{\textwidth}{\textheight}%
    }%
}

\vspace*{0.28cm}

\begin{center}
\begin{minipage}{0.94\textwidth}

{\color{TitleInk}
\centering
{\fontsize{24}{28}\selectfont
\bfseries
\itshape
\PaperTitle
\par}
}

\vspace{0.48cm}

{\fontsize{10.8}{13.0}\selectfont
\bfseries
\setlength{\parindent}{0pt}
\noindent
\PaperAuthors
\par
}

\vspace{0.36cm}

{\fontsize{8.9}{11.1}\selectfont
\setlength{\parindent}{0pt}
\noindent
$^{1}$College of Computing and Data Science, Nanyang Technological University, Singapore,
$^{2}$National University of Singapore, Singapore,
$^{3}$Alibaba-NTU Global e-Sustainability CorpLab (ANGEL), Nanyang Technological University, 637335, Singapore,
$^{4}$Shanghai Jiao Tong University, Shanghai, China,
$^{5}$State Key Laboratory for Development and Utilization of Forest Food Resources, Zhejiang A\&F University, Hangzhou, China,
$^{6}$University of Oxford, Oxford, United Kingdom,
$^{7}$Department of Medical Oncology, Shanghai East Hospital, Tongji University School of Medicine, Shanghai, China,
$^{8}$University of Southern California, Los Angeles, CA, USA,
$^{9}$China Zhejiang Key Laboratory of Intelligent Manufacturing for Functional Chemicals, ZJU-Hangzhou Global Scientific and Technological Innovation Center, Zhejiang University, Hangzhou, China,
$^{10}$Institute of Nanotechnology and Intelligence (inAI), College of Chemistry and Material Sciences, Jinan University, Guangzhou, China,
$^{11}$State Key Laboratory of Quantitative Synthetic Biology, Shenzhen Institute of Synthetic Biology, Shenzhen Institutes of Advanced Technology, Chinese Academy of Sciences, Shenzhen 518055, China
\par
}

\vspace{0.38cm}

{\fontsize{9.6}{11.7}\selectfont
\setlength{\parindent}{0pt}

\noindent
\textbf{\fontsize{11.0}{13}\selectfont Abstract: }
Protein language models organize sequence and structure at scale; here we add a global coordinate system for how proteins respond to mutation.\supercite{9,10,11,12,13,15,17,18,19,20,21,22}
We present \RegimeFormer{}, a large protein perturbation model coupled to \RegimeAtlas{}, which we constructed by harmonizing, representing and indexing 202,556,313 non-redundant protein sequences across the tree of life.
This full sequence universe defines the atlas-wide protein-regime map.
A diversity-preserving one-million-protein subset provides the high-resolution training and inference layer, with 995,995 proteins yielding 407,048,356 residue summaries and substitution-specific predictions available on demand.
Across experimental DMS, molecular benchmarks, structural confidence and evolutionary constraint, \RegimeFormer{} reveals reproducible protein-level perturbation regimes that organize residue fragility, adaptability and uncertainty.
Explicit regime conditioning improves substitution-specific reconstruction and shows its largest relative advantage under unseen-protein, unseen-family and low-homology evaluation.
The resulting molecular priors also improve downstream transcriptomic and drug-response modelling.
Together, \RegimeFormer{} and the 202.6-million-sequence \RegimeAtlas{} establish a large-model framework for mapping, predicting and querying protein perturbation landscapes across global sequence space.
\par
}

\vspace{0.24cm}

{\fontsize{9.3}{11.2}\selectfont
\setlength{\parindent}{0pt}


\par
}

\vspace{0.20cm}

{\fontsize{9.1}{11.0}\selectfont
\setlength{\parindent}{0pt}

\noindent
$^{\dagger}$ \textbf{These authors contributed equally to this work.}\\[2pt]

\noindent
$^{*}$ \textbf{Co-corresponding authors.}\\[2pt]

\noindent
\textbf{E-mail: }
\href{mailto:MASI0004@e.ntu.edu.sg}
{MASI0004@e.ntu.edu.sg}

\par
}

\end{minipage}
\end{center}

\end{titlepage}

\clearpage

Protein sequence space is vast, but the experimentally accessible fraction is narrow. Deep mutational scanning (DMS) and related multiplexed assays measure thousands of substitutions within a protein and have made it possible to observe sequence--function landscapes directly.\supercite{2,3,4,6,7,8} These measurements reveal sharply heterogeneous effects: some positions tolerate many amino-acid changes, others are highly constrained, and a smaller subset supports substitutions that improve activity, abundance, binding or fitness. Such maps have become central to variant interpretation and protein engineering, but each experiment is tied to a particular protein, readout and laboratory protocol. The resulting landscape collection remains sparse relative to the hundreds of millions of protein sequences now catalogued in public resources.\supercite{37,38,39}

Computational models have progressively reduced this coverage gap. Evolution-based approaches such as DeepSequence, EVE and GEMME use homologous sequence variation to infer the compatibility of amino-acid substitutions with a protein background, whereas protein language models learn sequence representations at much larger scale and can predict mutational effects without assay-specific training.\supercite{10,17,18,19,21} Structure-aware language models and AlphaFold-derived models add local geometric context, further improving variant-effect prediction in many settings.\supercite{15,20,33,34} These approaches establish powerful local predictors of substitution effect. At the scale of hundreds of millions of proteins, a higher-order question becomes accessible: whether entire mutational landscapes occupy reproducible global states, and whether those states explain why the same class of perturbation is tolerated in one protein but destabilizing or functionally disruptive in another.

Several lines of experimental and theoretical work make such organization plausible. Protein stability can buffer otherwise deleterious substitutions and increase evolvability,\supercite{23,28,29} pairwise and higher-order interactions make mutational effects dependent on sequence background,\supercite{24,25,26,27,30,31,32,45} and direct-coupling models show that residue-level constraints are structured by evolutionary interactions rather than by independent-site statistics.\supercite{45} At the same time, large-scale protein resources are already organized by family, domain and three-dimensional structure.\supercite{35,36,38,39} What is missing is a comparable coordinate system for perturbational behaviour itself: a representation that is defined at the protein level, manifests at the residue level, generalizes across families and taxa, and can be tested against experimental mutation maps.

We developed \RegimeFormer{} as the central model of the study and built \RegimeAtlas{} as its global sequence-space substrate. After harmonizing 424,786,370 source records, we retained 202,556,313 non-redundant proteins and computed the protein-level representation and regime index that define the atlas. From this full atlas we selected a diversity-preserving one-million-protein set for high-resolution RegimeFormer training and residue-scale inference. The model combines pretrained residue representations with protein-level regime coordinates, mutant identity, assay context and optional structural or evolutionary evidence, producing substitution effects, calibrated uncertainty and residue-level perturbation maps. We tested the learned regime space across representation backbones, taxa, families and data sources; mapped its residue-level functional organization; evaluated substitution-specific reconstruction against DMS; and challenged the model under sequence, temporal and assay shift. We then scaled high-resolution inference to 995,995 proteins and 407,048,356 residues, aligned the learned space with structural and evolutionary evidence, and carried RegimeFormer-derived molecular priors into transcriptomic and drug-response prediction.

\section*{RegimeFormer maps perturbation regimes across 202 million protein sequences}
RegimeAtlas was assembled from 424,786,370 raw sequence records obtained from UniProtKB, NCBI RefSeq, Ensembl/GENCODE, MGnify, environmental and metagenomic collections, and other public repositories. UniProtKB contributed 42.6\% of the raw aggregate, RefSeq 19.4\%, Ensembl/GENCODE 12.3\%, MGnify 8.2\%, environmental or metagenomic sources 6.8\%, and the remaining repositories 10.7\% (Supplementary Fig. 1). Sequence-level quality control removed invalid alphabets, very short fragments, low-complexity or highly ambiguous sequences and redundant records. Exact and near-duplicate collapsing followed by taxonomy reconciliation yielded 202,556,313 distinct proteins, corresponding to 47.7\% of the raw aggregate. These 202,556,313 sequences form the computational substrate of RegimeAtlas: every retained entry participates in the atlas-wide protein representation, regime-coordinate index and provenance layer used for global mapping, retrieval and cross-taxon organization. Taxonomy was resolved for 96.73\% of mappable entries, yielding broad coverage across bacteria, eukaryotes, archaea and viruses. Median protein length in the final atlas was 305 amino acids (interquartile range 167--512), with the expected shorter distribution for viral proteins and longer distribution for eukaryotic proteins (Supplementary Fig. 1).

We embedded the 202,556,313 atlas proteins with a pretrained ESM-family encoder and projected them into a continuous protein-regime space that captures perturbational state across taxonomic and family boundaries (Fig. 1a,b). The resulting atlas is therefore a protein-level perturbation map of the full harmonized sequence universe, while the smaller cohorts used later in the paper provide targeted experimental, structural or evolutionary validation. In the 24,918-protein subset with complete confounder annotations, adjusted mutual information between regime label and taxonomy was 0.0063 despite strong local taxonomic structure in the underlying PLM space. A classifier using PLM-derived features achieved regime balanced accuracy 0.599 and AUROC 0.865, whereas taxonomy, length and Pfam features together reached balanced accuracy 0.270. For the continuous coordinate, PLM-derived features explained 44.3\% of variance compared with 1.0\% for the conventional covariates, yielding a 0.435 increase in explained variance after addition of the learned representation (Fig. 1c). The atlas-wide coordinate thus captures a perturbational axis that generalizes across conventional protein groupings.

Two complementary control analyses quantify the regime signal at different levels. Figure 1d summarizes the absolute per-protein association, $|\rho|$, across downstream perturbational readouts. The real regime reaches mean $|\rho|=0.38$, exceeding length-only (0.30), shuffled embeddings (0.28), random tokens (0.27), amino-acid composition (0.29), species-only (0.29) and Pfam-only (0.28) by 0.08--0.11 (mean excess, 0.093). The positive 0.27--0.30 control floor follows directly from taking absolute correlations before aggregation. Supplementary Fig. 2 provides the signed matched-null analysis for mutation sensitivity: the real-regime correlation is 0.229 (95\% CI, 0.201--0.257), compared with 0.041 (0.018--0.064) after length matching, 0.036 (0.013--0.059) after length-plus-composition matching, and permutation means of 0.004 and 0.006 within species- and Pfam-matched sets. The absolute cross-readout summary and the signed matched-null test therefore converge on the same result: the atlas-derived regime carries perturbational signal beyond the tested nuisance representations. Data source explained 3.8\% of regime-coordinate variance, compared with 71.4\% attributed to the biological coordinate itself, and independent coordinate runs remained highly concordant (Spearman $\rho=0.991$; ICC=0.987), with alternative coordinate definitions correlated at $\rho=0.91$--0.96 (Supplementary Fig. 2).

The signal also transferred across representation families. ESM-2, ESM-1v and ProtT5 produced a mean cross-backbone regime correlation of 0.700, mean Cohen's $\kappa$ of 0.60 and 92\% agreement in the direction of biological effects (Extended Data Table 1). Functional annotation profiles varied systematically across the continuous landscape: fragile regimes were enriched for active and binding sites, enzyme annotations and Gene Ontology terms, whereas adaptive regions showed distinct functional compositions (Fig. 1e). These differences persisted when entire species, Pfam families or assays were held out. Relative to a length-only representation, mean absolute correlation increased by 0.41 in species hold-out, 0.48 in Pfam hold-out and 0.38 in assay hold-out, with all bootstrap intervals excluding zero (Fig. 1f). The combined evidence therefore supports a perturbational axis that cuts across conventional sequence categories while remaining biologically structured.

\clearpage
\begin{landscape}
\thispagestyle{plain}
\begin{center}
\makebox[\linewidth][c]{%
\includegraphics[
    width=1.05\linewidth,
    height=0.96\textheight,
    keepaspectratio
]{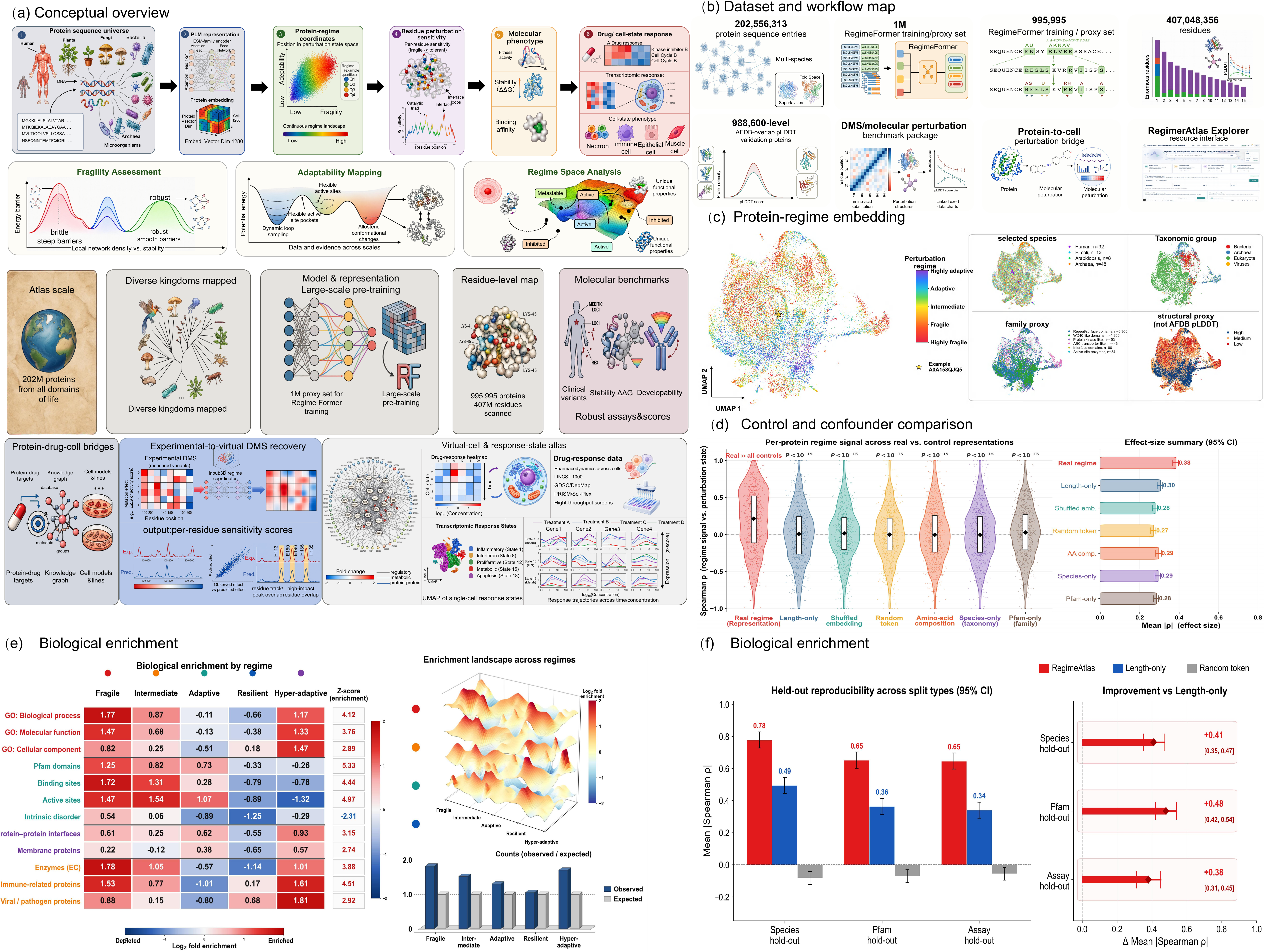}%
}
\end{center}
\end{landscape}
\clearpage
\captionsetup{font=small,justification=justified,singlelinecheck=false}
\captionof{figure}{\textbf{A perturbation-regime atlas of protein sequence space.} \textbf{a}, Conceptual framework linking the protein sequence universe, PLM representation, continuous regime coordinates, residue-level perturbation sensitivity, molecular phenotype and downstream cellular response. \textbf{b}, Dataset and workflow map separating the full 202,556,313-protein atlas, the one-million-protein high-resolution training/proxy subset, the 995,995-protein residue scan, molecular benchmarks, protein-to-cell bridge and Explorer. \textbf{c}, UMAP visualization of the protein-regime representation with selected species, taxonomic, family and structural overlays. The confounder audit uses 24,918 proteins; regime--taxonomy adjusted mutual information is 0.0063, while PLM features add 0.331 balanced accuracy beyond taxonomy, length and Pfam covariates. \textbf{d}, Control and confounder comparison using the mean absolute per-protein association, $|\rho|$. The real regime has mean $|\rho|=0.38$; length-only, shuffled-embedding, random-token, amino-acid-composition, species-only and Pfam-only controls have means of 0.30, 0.28, 0.27, 0.29, 0.29 and 0.28, respectively. The real-minus-control excess is 0.08--0.11. Error bars show 95\% CIs. The absolute-value statistic produces the observed positive control floor; Supplementary Fig. 2 provides the complementary signed matched-null analysis. \textbf{e}, Annotation enrichment across fragile, intermediate, adaptive, resilient and hyper-adaptive regime bins. \textbf{f}, Reproducibility under species-, Pfam- and assay-held-out evaluation. Relative to length-only representation, mean absolute Spearman correlation increases by 0.41, 0.48 and 0.38, respectively; intervals are 95\% bootstrap CIs.}
\clearpage
\section*{RegimeFormer resolves residue-level perturbation organization}
The atlas-wide protein regime resolves into a structured local landscape. We projected the continuous regime representation back onto individual sequences and summarized the predicted substitution distribution at each residue as three complementary quantities: fragility, representing the tendency for substitutions to be deleterious; adaptability, representing the potential for beneficial or tolerated substitutions; and predictive uncertainty (Fig. 2a,b). These tracks were heterogeneous within proteins. In EGFR, for example, high-fragility peaks concentrate within and around the kinase domain across the 1,210-residue sequence, whereas adaptable and high-uncertainty regions follow distinct local profiles (Fig. 2b). The global regime therefore acts as the protein background on which local perturbation sensitivity is organized.

We next asked whether the residue-level scores were concentrated at independently annotated functional sites. Across the reference-matched atlas-scale scan of 995,995 proteins and 407,048,356 residues, high-fragility residues were strongly enriched at active sites (odds ratio 11.3; 95\% CI, 7.6--16.8; $P=2.4\times10^{-187}$), ligand-binding sites (odds ratio 6.1; 4.3--8.7; $P=1.2\times10^{-145}$), protein--protein interfaces (odds ratio 3.2; 2.4--4.2; $P=4.7\times10^{-98}$), disease-associated missense positions (odds ratio 4.0; 3.2--4.9; $P=6.1\times10^{-112}$), post-translational modification sites (odds ratio 5.7; 4.2--7.1; $P=3.8\times10^{-156}$) and conserved motifs (odds ratio 2.8; 1.9--4.1; $P=5.3\times10^{-67}$) (Fig. 2c). The enrichment spans catalytic chemistry, ligand recognition, macromolecular interfaces, disease-associated positions and conserved sequence features, placing diverse functional constraints at the same end of the perturbational spectrum.

The association was continuous. Mutation sensitivity declined monotonically from fragile to adaptive regime states, with Pearson $r=-0.71$ and Spearman $\rho=-0.64$ in the binned analysis shown in Fig. 2d. The continuous coordinate resolves intermediate sensitivity states that are compressed by categorical bins. Downstream analyses remain stable after quantile or atlas-derived discretization (Fig. 5f), establishing the continuous regime score as the primary representation and the discrete labels as interpretable views of the same landscape.

Regime state also organized where existing predictors made errors. Across six established predictor families, mean absolute error was consistently highest in the fragile bin and declined toward the adaptive end of the spectrum (Fig. 2e). For ESM-650M, the fragile-to-adaptive error difference was 1.38; for EVE it was 1.23; for GEMME 1.13; for MSA Transformer 0.98; for a structure-aware baseline 0.87; and for the ensemble baseline 0.76. The same gradient across models trained with different objectives identifies regime position as a shared predictor-difficulty axis. A matched topology analysis further showed that topology-derived interaction features and regime-derived features were related but not interchangeable (Fig. 2f). Together, these results connect the protein-level coordinate to the local distribution of functional constraint and to the conditions under which mutation-effect prediction becomes difficult.

We next asked whether this difficulty axis could be predicted before the held-out experimental labels were revealed. A failure-forecasting model using assay and protein metadata alone achieved a Spearman correlation of 0.310 with observed predictor error; regime coordinates alone increased this to 0.380, and combining metadata with regime state increased it to 0.533 on held-out predictor families (Extended Data Fig. 4; Supplementary Table 18). The combined model classified high-error cases with AUROC 0.861. This analysis turns the fragile-to-adaptive error gradient into a forecastable quality-control variable that ranks where independent variant-effect predictors are most likely to incur large errors.

\clearpage
\thispagestyle{plain}

\begin{center}
\includegraphics[
    width=0.98\textwidth
]{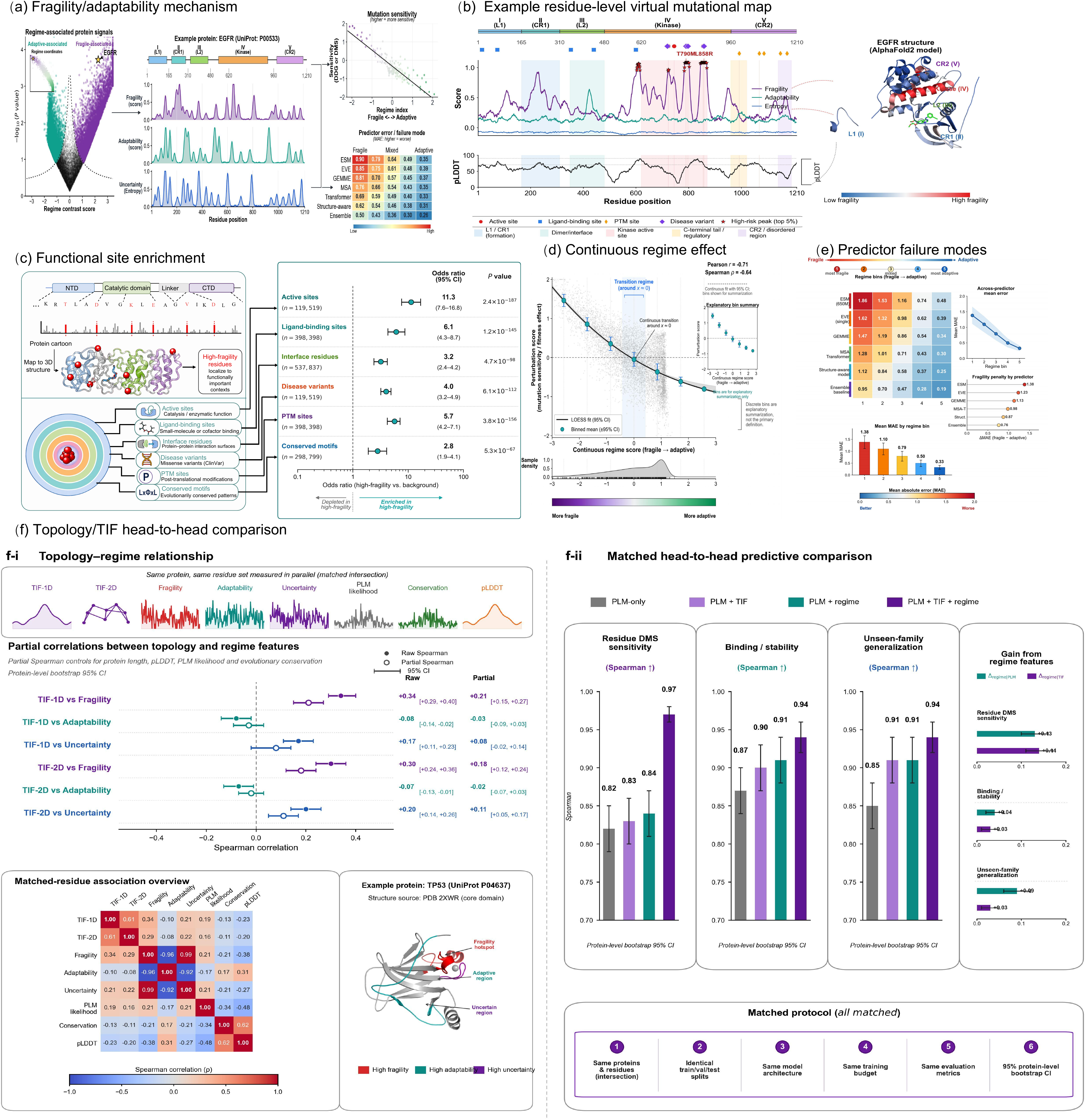}
\end{center}

\vspace{0.3em}

\captionsetup{
    font=small,
    justification=justified,
    singlelinecheck=false
}

\captionof{figure}{%
\textbf{Perturbation regimes organize residue-level sensitivity and model difficulty.}
\textbf{a}, Relationship between protein-level regime state and residue-level fragility, adaptability and uncertainty, with a representative sequence track.
\textbf{b}, EGFR virtual mutational map showing residue-level perturbation tracks, functional annotations and structural projection.
\textbf{c}, Enrichment of high-fragility residues at active sites, ligand-binding sites, interfaces, disease-associated positions, post-translational modification sites and conserved motifs. Points are odds ratios; horizontal bars are 95\% CIs.
\textbf{d}, Continuous relationship between regime score and mutation sensitivity (Pearson $r=-0.71$; Spearman $\rho=-0.64$ in the displayed analysis).
\textbf{e}, Mean absolute error of established variant-effect predictors across regime bins, ordered from fragile to adaptive.
\textbf{f}, Matched topology/TIF comparison, including partial association analysis and predictive combinations of PLM, topology and regime features.
}

\label{fig:regime-residue}

\clearpage
\section*{RegimeFormer reconstructs experimental mutational landscapes}
We next tested whether the protein-level perturbation regime supports full substitution-specific prediction in addition to residue ranking. Figure 3 separates the experimental bottleneck, the RegimeFormer operator, direct residue-level alignment with measured DMS, benchmark breadth, component ablations and strict out-of-distribution transfer. For each residue $i$ and candidate mutant amino acid $a$, RegimeFormer combines the residue-level PLM representation $h_i$, wild-type and mutant amino-acid embeddings, the protein-regime coordinate $z_{\mathrm{regime}}$, assay context $q_{\mathrm{assay}}$ and optional structural or evolutionary covariates. Regime-aware attention and assay conditioning feed substitution-effect, uncertainty and three-class effect heads, after which substitution-level outputs are aggregated into residue-level fragility, adaptability and uncertainty summaries (Fig. 3a,b).

The first test was whether these residue-level summaries aligned with experimental landscapes before any aggregate benchmark was considered. In the representative EGFR scan, predicted fragility follows the high-sensitivity peaks of the experimental DMS track, while predicted adaptability and uncertainty identify distinct regions of tolerated or heterogeneous substitution response (Fig. 3c). Across 186 DMS proteins, median per-protein Spearman agreement was 0.53 for fragility versus deleterious-substitution fraction, 0.52 for adaptability versus tolerated or beneficial-substitution fraction, and 0.46 for uncertainty versus substitution-effect variance; macro-averages were 0.54, 0.52 and 0.47, respectively. The top-ranked residues were substantially enriched for the expected experimental phenotype: the top 5\% fragile residues showed 21.2-fold enrichment for deleterious substitutions, the corresponding high-effect deleterious set showed 17.0-fold enrichment, and the top 5\% adaptable residues showed 16.3-fold and 15.0-fold enrichment for tolerated and beneficial substitutions, respectively. High-uncertainty residues also concentrated experimental effect variance, linking the uncertainty head to a measurable property of the mutational landscape rather than to a post hoc confidence score.

We next expanded the benchmark across molecular tasks. Across 217 substitution assays, RegimeFormer reached aggregate Spearman $\rho=0.698$, essentially matching the highest displayed supervised aggregate, AUTOSCIENTISTS ($\rho=0.700$), while exceeding Kermut ($0.662$), ProteinNPT ($0.613$) and the displayed embedding and zero-shot baselines (Fig. 3d; Supplementary Fig. 5). Task-stratified performance remained high across activity ($\rho\approx0.65$), binding ($\rho\approx0.67$), expression ($\rho\approx0.70$), fitness ($\rho\approx0.62$) and stability ($\rho\approx0.85$). The zero-shot panel provides the complementary unsupervised operating regime, while the supervised comparison tests the full substitution-specific model. Across both views, RegimeFormer remains strong across heterogeneous molecular readouts and assay scales.

Substitution-specific ablations then separated the contribution of each input. The full model reached a substitution-level Spearman correlation of 0.723, RMSE of 0.785, damaging-mutation AUPRC of 0.883 and residue-aggregated Spearman correlation of 0.841 (Fig. 3e). Removing the mutant-amino-acid embedding produced the largest biologically meaningful drop, reducing substitution Spearman to 0.548; an ESM/PLM-only model reached 0.579. Removing the wild-type embedding reduced correlation to 0.657, removing assay conditioning to 0.693 and removing regime coordinates to 0.701. Shuffling substitution labels collapsed correlation to 0.281 and increased RMSE to 1.098. The ordering of these ablations is important: mutant identity carries the largest local signal, whereas the protein regime acts as a contextual prior that improves the interpretation of the same substitution in different protein backgrounds.

The same model was then evaluated under progressively harder sequence shift. On strict unseen-protein, unseen-family and low-homology ($<30\%$ identity) splits, the full model retained approximately 82\%, 62\% and 38\% of its in-distribution substitution-level performance, respectively, and remained the highest-performing model across the displayed substitution-level, classification and residue-aggregated metrics (Fig. 3f). The sequence-identity curves make the same point continuously: all methods improve as similarity to the training set increases, but the regime-aware model separates most clearly from PLM-only and topology baselines in the low-homology range. This behaviour is consistent with the regime coordinate providing protein-level context that becomes more valuable when local sequence neighbourhoods are sparse. Cross-species and cross-laboratory evaluations provide complementary shifts: species-held-out performance retained 92.3\% of the in-distribution level across eight species (Supplementary Fig. 10), and cross-laboratory transfer retained a median 85\% of within-laboratory performance for the regime-aware model (Extended Data Table 2).

Calibration and reproducibility were evaluated separately from rank performance. Supplementary Fig. 3 shows that the uncertainty head reduces ECE from 0.094 to 0.024 and NLL from 0.742 to 0.291, while five-seed and hyperparameter analyses remain in the same performance basin. Selective prediction provides an operational test of this calibration: restricting to progressively lower-uncertainty subsets increases selective Spearman from 0.548 at full coverage to 0.671 at 60\% coverage and 0.771 at 20\% coverage (Supplementary Table 7). On replicated assays, the median model-to-experiment agreement reaches 91.1\% of replicate-to-replicate agreement (Supplementary Table 4), and the full model recovers 61.2\% of score variance compared with 40.2\% for PLM-only prediction (Supplementary Table 6).

Supplementary Fig. 5 decomposes the same expanded benchmark rather than introducing a separate headline dataset. Nesting the assay-level results by UniProt protein yields 87 proteins for a direct Kermut comparison: RegimeFormer reaches mean Spearman $\rho=0.698$ versus $0.662$ for Kermut, wins on 81 of 87 proteins and gives a mean paired gain of $+0.036$ (95\% CI, 0.031--0.040). The separation is largest in the smallest assay quintile and narrows as experimental coverage increases, consistent with the regime prior contributing most when assay-specific supervision is sparse. Extended Data Fig. 1 complements this population-level analysis with explicit avGFP, BLAT, PABP and HRAS mutation matrices, and the frozen post-cutoff challenge shows little temporal degradation: mean Spearman changes from 0.591 internally to 0.585 on later-release data ($\Delta\rho=-0.006$, 95\% CI, $-0.020$ to 0.007; Extended Data Fig. 2; Supplementary Table 11). Together, these analyses connect residue-level landscape reconstruction, broad molecular benchmarking and temporal transfer within a single substitution-specific framework.

\clearpage
\begin{landscape}
\thispagestyle{plain}
\begin{center}
\includegraphics[
    width=0.99\linewidth,
    height=0.94\textheight,
    keepaspectratio
]{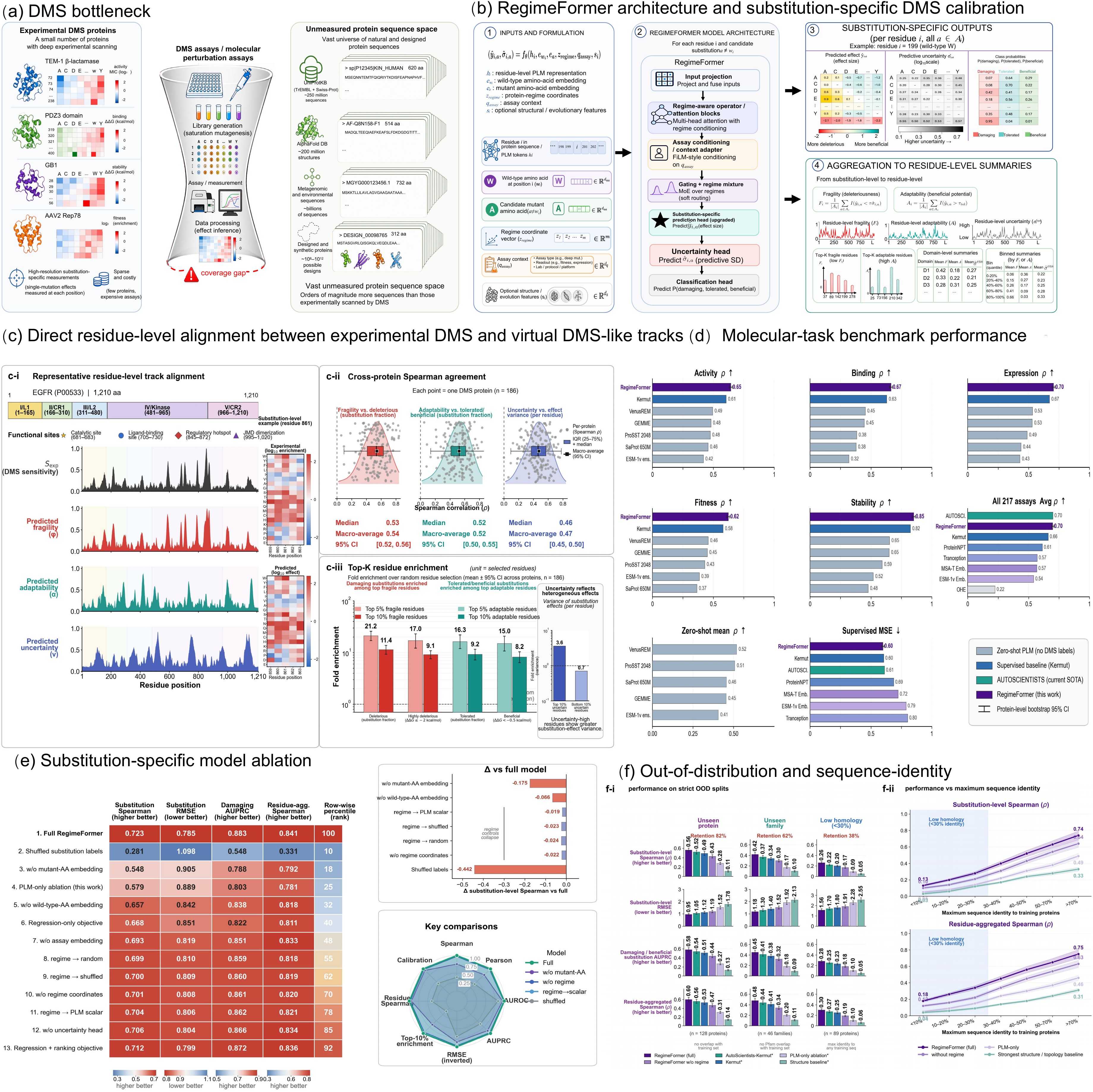}
\end{center}
\end{landscape}
\clearpage

\captionsetup{
    font=small,
    justification=justified,
    singlelinecheck=false
}

\captionof{figure}{\textbf{RegimeFormer reconstructs experimental substitution effects across proteins and molecular tasks.} \textbf{a}, Experimental DMS densely measures single-substitution effects for a small set of proteins, leaving a large gap relative to natural, metagenomic and designed sequence space. \textbf{b}, RegimeFormer formulation, architecture, substitution-specific outputs and aggregation to residue-level fragility, adaptability and uncertainty. Inputs include residue PLM features, wild-type and mutant amino-acid embeddings, protein-regime coordinates, assay context and optional structure/evolution features. \textbf{c}, Direct residue-level alignment with experimental DMS. In 186 proteins, median per-protein Spearman agreement is 0.53 for fragility, 0.52 for adaptability and 0.46 for uncertainty; top-ranked fragile and adaptable residues are strongly enriched for the corresponding experimental substitution classes. \textbf{d}, Molecular-task benchmark across 217 substitution assays. RegimeFormer reaches aggregate Spearman $\rho=0.698$, within 0.002 of the highest displayed supervised aggregate, and remains competitive across activity, binding, expression, fitness and stability. \textbf{e}, Substitution-specific ablation. The full model reaches Spearman 0.723, RMSE 0.785, damaging AUPRC 0.883 and residue-aggregated Spearman 0.841; removing mutant-amino-acid identity or reducing the model to PLM-only prediction causes the largest non-random degradations. \textbf{f}, Strict out-of-distribution evaluation on unseen proteins, unseen families and low-homology proteins, together with performance as a function of maximum sequence identity to training proteins. The full model retains the strongest performance as sequence similarity decreases. Error bars and shaded regions denote protein-level bootstrap uncertainty where shown.}

\clearpage
\section*{RegimeFormer scales virtual mutational maps across 407 million residues}
RegimeAtlas uses a two-resolution architecture. The 202,556,313-protein atlas supplies the global protein-level regime map, and a diversity-preserving one-million-protein subset supplies high-resolution residue and substitution modelling. Of these one million proteins, 995,995 passed the residue-scale pipeline and yielded summaries for 407,048,356 residues (Fig. 4a,b). For each residue we stored fragility, adaptability and uncertainty together with top-ranked sites and ten-bin or domain-level summaries, while the same model serves full $L\times19$ substitution predictions for focused queries. This design couples global sequence-space coverage to residue-level resolution without reducing the atlas itself to the one-million-protein scan.

This streaming design preserved numerical fidelity. Across the production shards, 95.1\% completed on the first pass and another 4.4\% were recovered after retry; 93.6\% of retried shards were successfully recovered (Supplementary Fig. 4). Residue-summary missingness was below 0.17\% for all major stored quantities across the full atlas and remained below 0.71\% even in the least complete unclassified subset. Top-ranked fragile residues were stable across seeds, with mean Jaccard overlap of 0.956 for the top 1\%, 0.972 for the top 5\% and 0.983 for the top 10\%; adaptable residues showed similarly high overlap. Streaming ten-bin profiles agreed with reference calculations at Spearman $\rho=0.998$, and domain-level summaries at $\rho=0.995$. Residue-level fragility, adaptability and uncertainty had streaming-to-reference correlations of 0.997, 0.996 and 0.994, respectively.

The storage reduction was substantial. Relative to materializing every substitution matrix, residue summaries alone required 0.78\% of the storage, summaries plus top-$K$ sites 2.3\%, and the deployed summaries plus on-demand index 0.35\% (Supplementary Fig. 4). On-demand substitution queries reproduced the offline reference with Pearson and Spearman correlations of 0.997, RMSE of 0.110 and uncertainty ECE of 0.014. Latency was benchmarked separately for the atlas inference engine, the version-pinned mutation-query replay and the deployed HTTP endpoints, because these operations include different amounts of computation and serving overhead; their timings are therefore reported with explicit definitions in Supplementary Figs. 4 and 10 and Supplementary Table 21 rather than collapsed into one ``single-query'' number. The main processing limitation was sequence length: success remained above 99\% through 1,000 residues but fell to 83.4\% for sequences longer than 10,000 residues, where timeouts and retry load increased (Supplementary Fig. 4).

The atlas summaries were then evaluated as an experimental prioritization layer. Fragile and adaptable residues were non-randomly distributed across functional sites and protein regions (Fig. 4c,d), and fixed-budget retrospective replay showed that ranking by \RegimeFormer{} recovered beneficial variants more efficiently than the strongest baseline. At a 5\% candidate budget, HitRate was 0.540 for \RegimeFormer{} and 0.434 for the strongest baseline, a 24\% relative gain; the advantage persisted from 1\% through 40\% budgets (Extended Data Fig. 3; Supplementary Table 17). In a separate blinded fixed-budget utility analysis, experts given regime coordinates, virtual DMS maps and uncertainty achieved HitRate@20 of 0.41 compared with 0.29 using sequence, annotations and a baseline predictor alone; Recall@20 increased from 0.27 to 0.38 and regret fell from 0.26 to 0.14 (Supplementary Table 19).

Large-scale summaries were also checked against a wild-type proxy reference. Of the 995,995 processed proteins, 987,412 (99.14\%) matched the proxy-reference records, and rank agreement across the resulting fragility, adaptability and uncertainty summaries was between 0.9991 and 0.99997 (Fig. 4e). Figure 4f illustrates how the same representation nominates a compact set of residues for saturation mutagenesis, biochemical or functional assays and orthogonal structural validation. RegimeAtlas therefore turns a 202.6-million-protein map into an experimentally actionable hierarchy: global protein coordinates for navigation, residue summaries for prioritization and substitution-specific predictions for selected targets.

\clearpage
\thispagestyle{plain}

\begin{center}
\makebox[\textwidth][c]{%
\includegraphics[
    width=0.98\textwidth,
    height=0.88\textheight,
    keepaspectratio
]{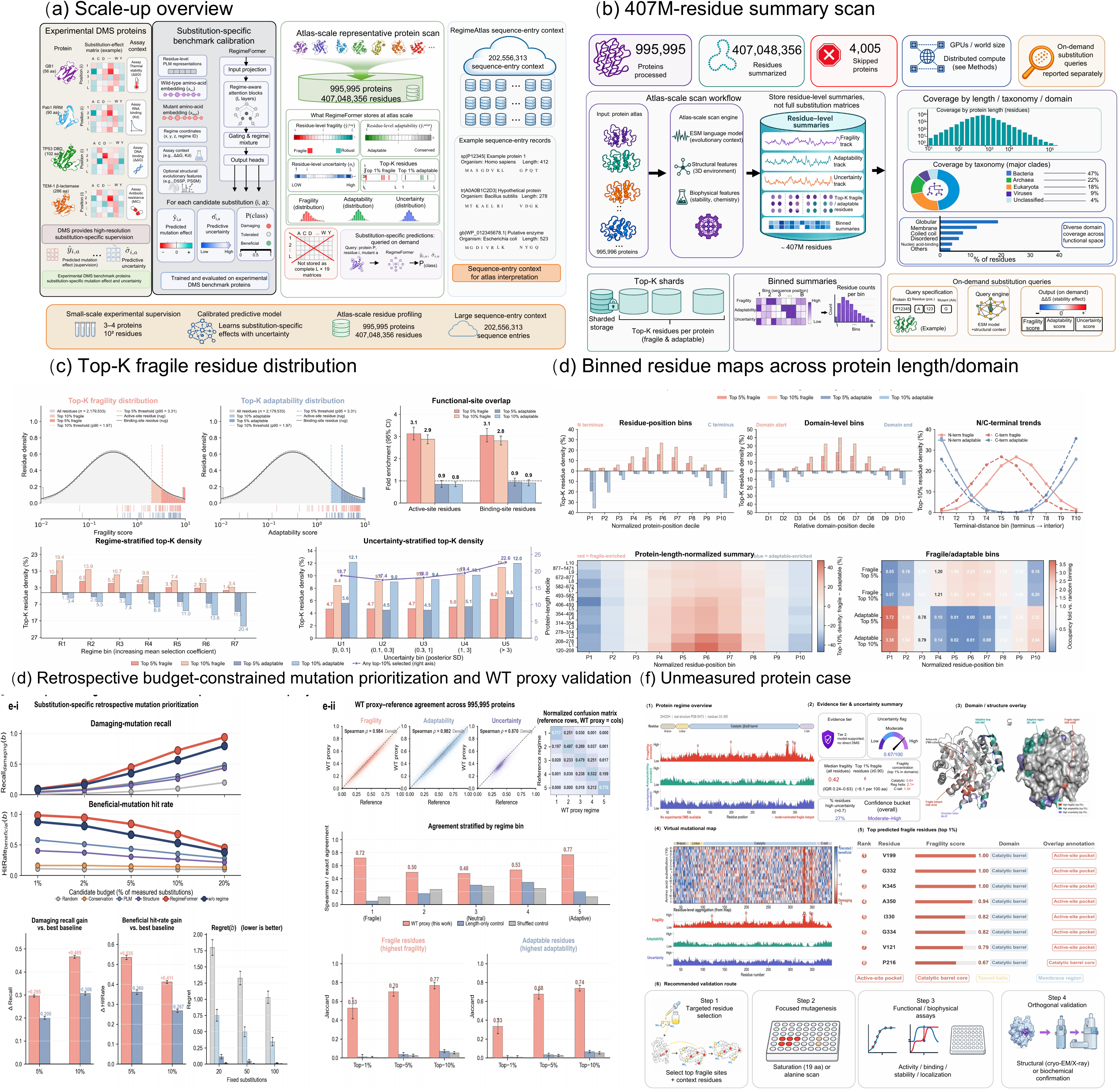}%
}
\end{center}

\vspace{0.3em}

\captionsetup{
    font=small,
    justification=justified,
    singlelinecheck=false
}

\captionof{figure}{\textbf{Atlas-scale virtual mutational maps and experimental prioritization.} \textbf{a}, Scale-up from experimentally supervised proteins to the full 202,556,313-protein atlas to the one-million-protein high-resolution scan. \textbf{b}, Streaming workflow used to process 995,995 proteins and summarize 407,048,356 residues. Stored outputs are residue-level fragility, adaptability, uncertainty, top-$K$ sites and binned or domain-level summaries; full substitution matrices are served on demand for focused targets. \textbf{c}, Distribution of top-ranked fragile and adaptable residues and their enrichment at functional sites. \textbf{d}, Position-normalized and domain-normalized maps showing where top-ranked residues occur across proteins of different lengths. \textbf{e}, Retrospective fixed-budget prioritization and wild-type-proxy reference agreement. \textbf{f}, Representative unmeasured-protein case showing virtual mutational tracks, evidence summary, top predicted fragile sites and a focused experimental validation route.}

\clearpage
\section*{RegimeFormer aligns with structural and evolutionary constraint}
We next asked whether the perturbation state learned from sequence and DMS supervision aligned with independent manifestations of protein constraint. The structural analysis intersected the large RegimeFormer scan with AlphaFold DB and retained 988,600 AFDB-overlap proteins after sequence matching and quality control (Fig. 5a; Supplementary Fig. 9). Rather than reducing structure to a single binary annotation, the new analysis organizes these proteins into five pLDDT-supported evidence tiers spanning fragile-supported, fragility--uncertainty-supported, uncertainty-supported, adaptability--stability-supported and adaptive-supported states. The tiers contain 98,860, 197,720, 271,865, 242,207 and 177,948 proteins, respectively, thereby covering the full structural-validation set while preserving a continuous transition from low- to high-confidence perturbation evidence.

Across all 988,600 proteins, local regime evidence correlated with pLDDT at Spearman $\rho=0.375$ (95\% CI, 0.368--0.381). The component relationships were directional: fragility evidence was negatively associated with pLDDT ($\rho=-0.428$), adaptability positively associated ($\rho=0.452$), and uncertainty negatively associated ($\rho=-0.188$) (Fig. 5a). The quartile analysis resolved the same trend in enrichment space. Fragile-supported proteins were enriched 2.02-fold in the lowest pLDDT quartile and depleted to 0.45-fold in the highest quartile, whereas adaptive-supported proteins were depleted to 0.64-fold in the lowest quartile and enriched 1.56-fold in the highest (Fig. 5b). Intermediate evidence tiers shifted progressively between these extremes, while a shuffled-regime control remained near the expected enrichment of 1.0 across quartiles. The association also survived density-level controls: within-protein shuffling reduced the correlation to 0.032, a length-only control to 0.064 and a global-mean pLDDT control to 0.089 (Fig. 5c). Thus, structural alignment follows the learned perturbation coordinate rather than a single protein-size or confidence surrogate.

Evolution provided a second independent axis. Across 57,048 phyloP-matched positions, conservation increased monotonically from fragile to adaptive RegimeFormer bins, with Spearman $\rho=0.5895$ (Fig. 5d). Of these positions, 51,872 (90.9\%) had CDS-ready codon records and 46,118 (80.9\%) remained after strict one-to-one ortholog filtering; the retained ortholog depth had a median of 18 species (IQR, 9--35), leaving 10,930 positions unmatched after the strict pipeline. Codon-level evidence converged with the residue-level trend: normalized phyloP support increased from 0.30 in fragile proteins to 0.82 in adaptive proteins, codon-level $\omega$-based support from 0.35 to 0.74, $d_N-d_S$ support from 0.33 to 0.78, and strict-ortholog support from 0.42 to 0.80 (Fig. 5d). Alternative $d_N/d_S$ estimators and ortholog-depth bootstraps gave concordant results (Supplementary Fig. 9), indicating that the evolutionary relation is not tied to a single codon model.

The perturbation coordinate also retained structure across major evolutionary groups. In the full cross-kingdom distributions, median regime score was $-0.37$ in bacteria, $+0.20$ in archaea, $+0.45$ in eukaryotes and $+0.83$ in viruses; within eukaryotes, the corresponding medians were $+0.49$ for fungi, $+0.33$ for plants and $+0.28$ for animals (Fig. 5e). A balanced visualization subset retained 540,000 bacterial, 72,000 archaeal, 180,000 eukaryotic, 38,000 viral, 55,000 fungal, 46,000 plant and 69,000 animal proteins so that bacterial abundance did not dominate the comparison. These kingdom-level shifts were not used to define the regime and coexisted with broad within-kingdom distributions, supporting a cross-species coordinate rather than a taxonomic label. Species-held-out and low-homology DMS analyses provide the corresponding predictive test (Supplementary Fig. 10).

Finally, the main conclusions were stable to how the continuous coordinate was discretized or supported. Across quantile resolutions from Q3 to Q10, pairwise agreement with the continuous reference remained high across residue sensitivity, functional enrichment, molecular benchmarking, cross-species consistency and structural consistency, with direction consistency typically above 90\% (Fig. 5f). Atlas-derived, wild-type-proxy, pLDDT-supported, entropy-supported and hydrophobicity/structure-supported definitions remained mutually concordant, whereas length-only and shuffled controls separated sharply from the biological definitions. In the explicit control-separation analysis, the atlas-derived definition reached a $z$ score of 6.4, compared with 5.2 for the wild-type proxy, 4.7 for pLDDT-supported, 4.2 for entropy-supported and 4.0 for hydrophobicity/structure-supported definitions, but only 0.2 for the length-only control and 0.0 after shuffling (Fig. 5f). Together, the structural, evolutionary and cross-kingdom analyses place RegimeFormer in established biological coordinates without collapsing the perturbation regime into any one of them.

\clearpage
\begin{landscape}
\thispagestyle{plain}
\begin{center}
\includegraphics[
    width=0.99\linewidth,
    height=0.94\textheight,
    keepaspectratio
]{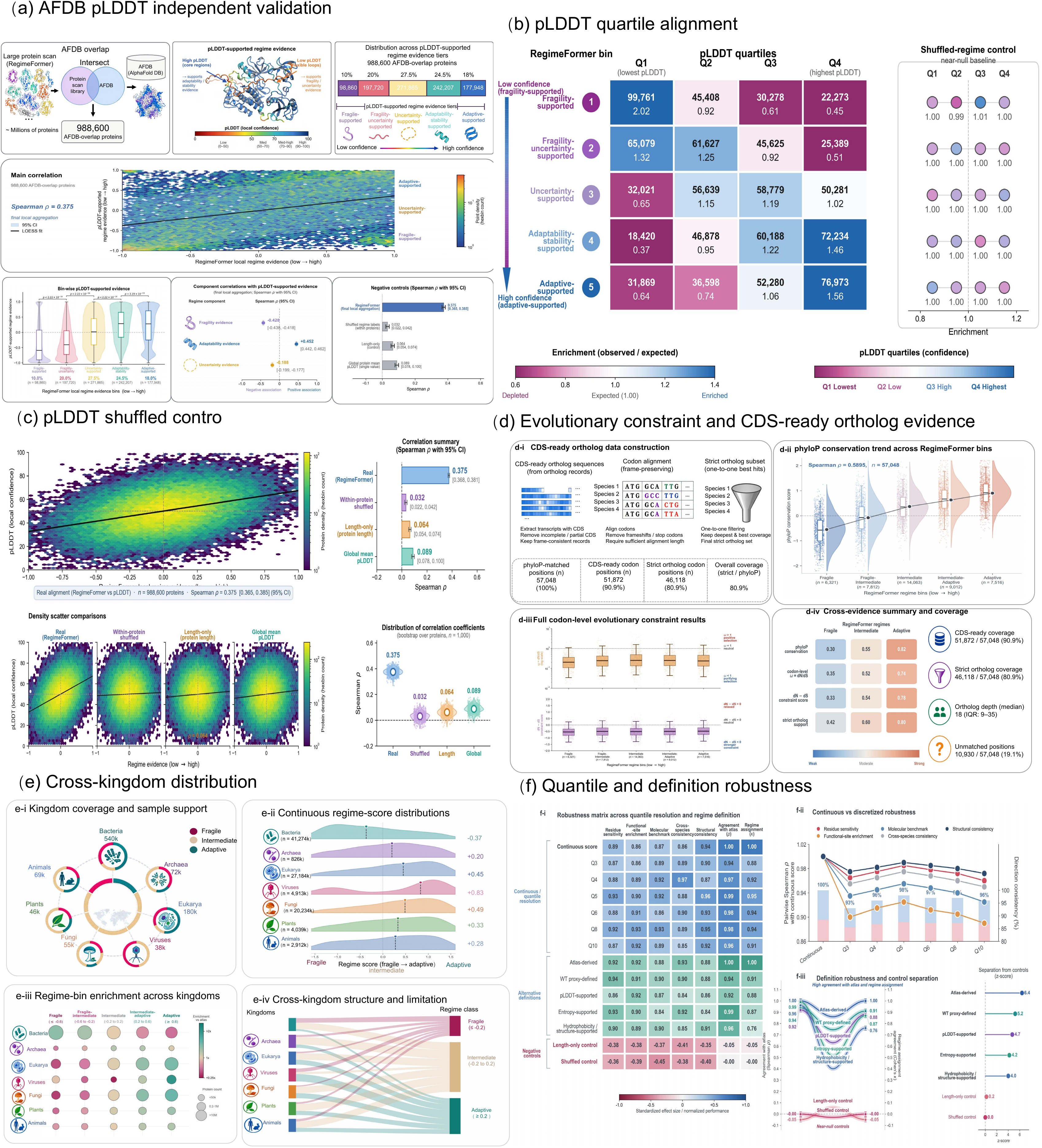}
\end{center}
\end{landscape}
\clearpage

\captionsetup{
    font=small,
    justification=justified,
    singlelinecheck=false
}

\captionof{figure}{\textbf{Independent structural, evolutionary and cross-kingdom validation of the learned perturbation state.} \textbf{a}, AlphaFold DB intersection and pLDDT-based validation in 988,600 quality-controlled overlap proteins. Five pLDDT-supported regime-evidence tiers span fragile-supported to adaptive-supported states; the main density analysis gives Spearman $\rho=0.375$ between local regime evidence and pLDDT, with component correlations shown for fragility, adaptability and uncertainty. \textbf{b}, Alignment of the five evidence tiers with pLDDT quartiles. Fragile-supported proteins are enriched in the lowest-confidence quartile, whereas adaptive-supported proteins are enriched in the highest; the shuffled-regime control remains near the null expectation. \textbf{c}, Density-level pLDDT association and negative controls based on within-protein shuffling, protein length and global-mean pLDDT. \textbf{d}, Evolutionary validation combining CDS-ready ortholog construction, the phyloP conservation trend across RegimeFormer bins, codon-level constraint and a cross-evidence coverage summary. Of 57,048 phyloP-matched positions, 51,872 are CDS ready and 46,118 pass strict orthology filters. \textbf{e}, Cross-kingdom support, continuous regime-score distributions, regime-bin enrichment and the mapping between kingdom and regime class. \textbf{f}, Robustness across quantile resolution and alternative regime definitions, including pairwise agreement, direction consistency and explicit separation from length-only and shuffled controls.}

\clearpage
\section*{RegimeFormer bridges protein perturbations to cellular responses}
We extended the protein perturbation map into cellular response through a three-layer data architecture (Fig. 6a). The portal-wide resource layer contains 19,824 proteins, 8,613 drugs and 412 cell types together with pathways and tissues (Supplementary Fig. 7a,b). Its cellular-response corpus comprises 3,176,765 upstream profiles across SciPlex3, Norman, Dixit, LINCS and Tahoe, with dataset coverage, perturbation composition, split manifests, residuals, calibration and top-50 DEG recovery reported in Supplementary Fig. 7c--h. Feature-complete matching then yields the task-specific Fig. 6 benchmark bridge: 467,150 records connecting 4,137 nodes through 269,337 edges, including 477 proteins, 38 drugs, 563 cell lines and 93 genes. Protein, drug and cell-line matching rates within this bridge are 94.2\%, 89.5\% and 97.1\%, respectively. This hierarchy links the broad portal graph to the response corpus and finally to the benchmark object used for quantitative protein-to-cell evaluation.

On transcriptomic perturbation data, \RegimeFormer{} carried protein-level information into heterogeneous cellular contexts. In SciPlex3, the regime-aware model achieved RMSE 0.0421 and Pearson correlation 0.294, compared with 0.0422 and 0.253 for the txPert/GEARS comparator; top-50 differentially expressed gene Jaccard was 0.863 versus 0.871 (Fig. 6b). On Norman 2019, RegimeFormer reached RMSE 0.124, Pearson 0.549 and DEG Jaccard 0.324; on Dixit 2016 the corresponding values were 0.129, 0.641 and 0.574. The basal-only control was substantially weaker in SciPlex3 (RMSE 0.337; Pearson 0.069; DEG Jaccard 0.031). Together, these results show that the protein-regime representation transfers measurable signal into cell-state prediction across distinct perturbation datasets.\supercite{55,57,58,59,60}

Drug response provided a stronger distribution-shift test. On PRISM 25Q2, comprising 1,101,954 rows from 727 cell lines and 1,482 compounds, random-split prediction reached $R^2=0.833$ and Pearson $r=0.913$; holding out entire cell lines retained $R^2=0.746$ and $r=0.866$ (Fig. 6c). Holding out entire compounds produced $R^2=0.038$ and $r=0.222$, defining chemical novelty as the hardest generalization regime. Performance followed compound similarity continuously: $R^2$ was 0.42 for Tanimoto similarity above 0.5, 0.12 for similarity 0.3--0.5 and 0.04 below 0.3, while ECE increased from 0.031 on the random split to 0.182 for unseen compounds (Supplementary Fig. 8). The contrast between strong unseen-cell transfer and similarity-dependent unseen-compound transfer directly identifies the additional representation needed for distant chemical space.\supercite{59,60,62}

Ablation across the protein-to-cell bridge showed that adding regime and virtual-map information improved the full representation over context and PLM features alone (Fig. 6d). In the representative A431--EGFR gefitinib case, inclusion of the virtual mutational map increased expression-change Pearson correlation from 0.54 to 0.87 and pathway recovery from 0.44 to 0.78, while drug-response RMSE decreased from 0.42 to 0.18 and top-50 DEG Jaccard increased from 0.47 to 0.82 (Fig. 6d). The same section of the resource exposes pathway-level transcriptional programs, cell-state embeddings and uncertainty so that a protein-level prediction can be inspected in its cellular context rather than returned as an isolated scalar.

We finally exposed the atlas through the \RegimeAtlas{} Explorer, using the model outputs as a common evidence layer spanning protein, residue, structural and cellular scales (Fig. 6e). The platform overview is shown in Extended Data Fig. 5: the landing surface links the Assistant, Protein Explorer, Virtual DMS, cell-state, communication, drug-response and benchmark modules, whereas the paired methods/provenance view exposes the full 202,556,313-protein atlas, the coordinate-analysis subset, release version and contributing sequence, structure and perturbation resources. Placing these two views together makes the intended interaction explicit: discovery-oriented interfaces and provenance-oriented interfaces are different views of the same frozen evidence object within one versioned application.

At the protein level, natural-language queries resolve to structured mutation records rather than free-form text alone. For EGFR, the Assistant returns ranked residue positions, sensitivity scores and the substitutions predicted to be most deleterious (Extended Data Fig. 6); for TP53, the corresponding Assistant view summarizes the protein-wide sensitivity profile and returns a table of high-scoring hotspots together with their mapped functional context (Supplementary Fig. 11). The structural workspace projects the same TP53 virtual-DMS object onto the three-dimensional protein model and records the active structure object and scene history (Supplementary Fig. 12). These views expose one perturbational object at residue, tabular and structural resolution. The DMS Studio supplies the complementary matrix representation: for TP53 (P04637), the interface renders the 20-amino-acid substitution matrix over the selected sequence window while retaining the protein-level sensitivity summary (Extended Data Fig. 8).

The same evidence architecture is used for the protein-to-cell bridge. An EGFR-inhibition query returns a predicted fitness effect of 0.53 with a 95\% interval of 0.27--0.79 together with uncertainty and the evidence classes used by the response model (Extended Data Fig. 7). The view exposes cell-response prediction, uncertainty and provenance as one traceable object. Backend audits are reported in Supplementary Fig. 10 and Supplementary Tables 20 and 21. A pinned-version reproducibility test gave Spearman 0.998 and MAE 0.004 relative to the offline reference, with 0.99 top-$K$ overlap and identical responses under the same version identifier. In the grounding audit, the full research assistant reached task accuracy 0.90, citation precision 0.96, unsupported-claim rate 0.04, tool-call success 0.97 and run-to-run ICC 0.91. The spatial communication module provides a further demonstration using a human lymph-node Visium dataset, where 4,035 spatial spots and seven cell types were linked through curated ligand--receptor pairs including CXCL12--CXCR4, CCL21--CCR7 and CCL19--CCR7 (Fig. 6f; Supplementary Fig. 10). The five core interface screenshots in Extended Data Figs. 5--8, together with the detailed TP53 Assistant and structural views in Supplementary Figs. 11 and 12, trace the same versioned RegimeFormer evidence object from atlas navigation to residue ranking, substitution matrices, three-dimensional interpretation and cellular-response queries.

\clearpage
\thispagestyle{plain}

\begin{center}
\makebox[\textwidth][c]{%
\includegraphics[
    width=0.98\textwidth,
    height=0.88\textheight,
    keepaspectratio
]{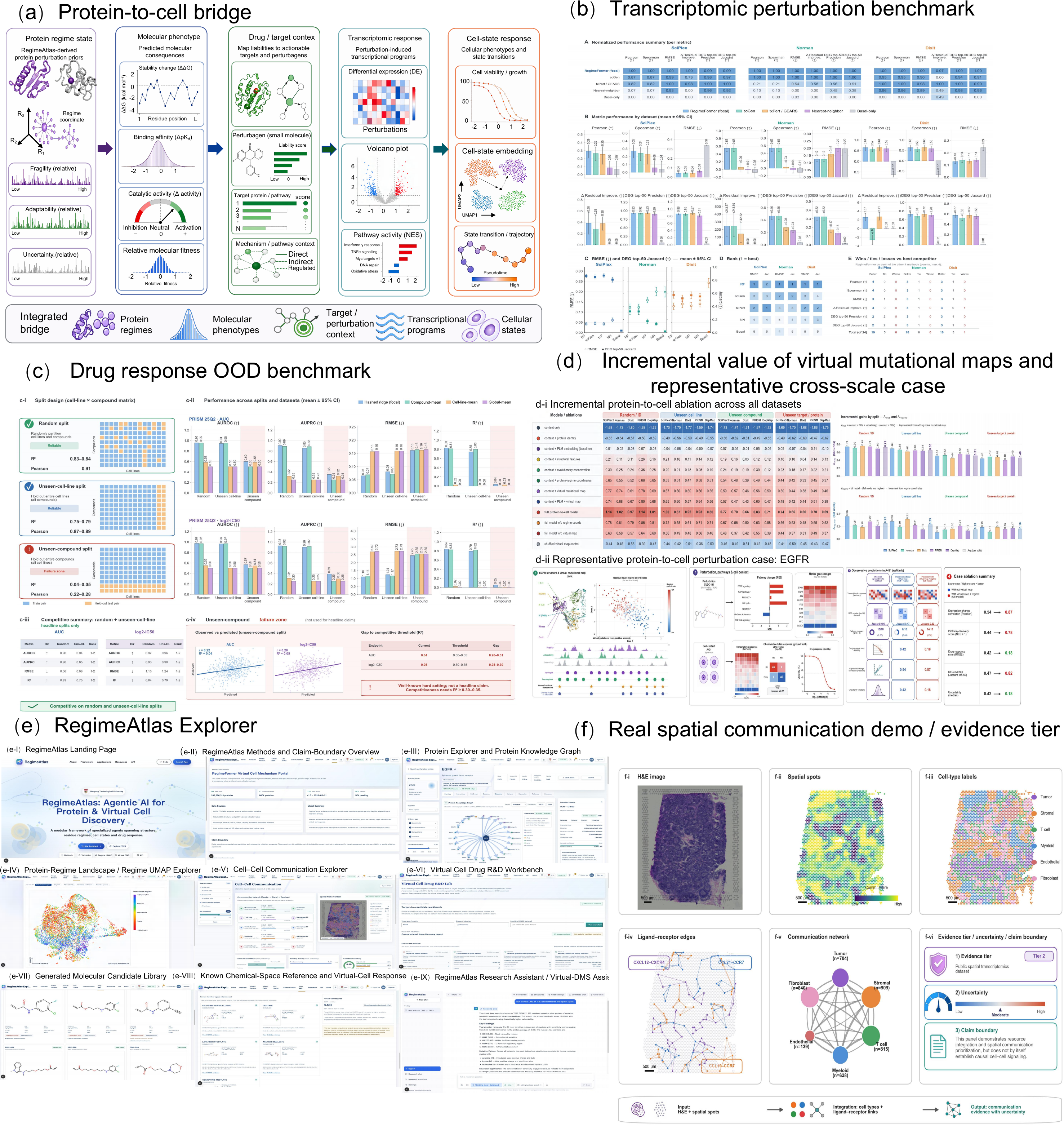}%
}
\end{center}

\vspace{0.3em}

\captionsetup{
    font=small,
    justification=justified,
    singlelinecheck=false
}

\captionof{figure}{\textbf{Protein perturbation information bridges molecular and cellular responses.} \textbf{a}, Task-specific protein-to-cell benchmark bridge linking protein regime state and molecular phenotypes to perturbagen context, transcriptional programs and cell-line response. The broader portal-wide knowledge graph is reported separately in Supplementary Fig. 7. \textbf{b}, Transcriptomic perturbation benchmarks on SciPlex3, Norman and Dixit datasets, including RMSE, Pearson correlation and top-50 DEG overlap. \textbf{c}, PRISM drug-response evaluation under random, unseen-cell-line and unseen-compound splits. The unseen-compound split falls into a distinct failure regime ($R^2=0.038$, Pearson $r=0.222$). \textbf{d}, Incremental protein-to-cell ablation and representative EGFR/gefitinib case. \textbf{e}, RegimeAtlas Explorer modules for protein search, regime navigation, virtual DMS, cell-state and drug-response analysis. \textbf{f}, Spatial communication demonstration in human lymph node, including spot-level cell states, ligand--receptor edges, network summary and evidence tier.}

\clearpage
\section*{Discussion}
The central result of this study is a 202,556,313-protein perturbation atlas in which mutational landscapes occupy reproducible protein-level regimes. The full harmonized sequence universe is the substrate of RegimeAtlas: protein-level representations and regime coordinates organize global sequence space, while nested experimental and structural cohorts test specific properties of that atlas. The one-million-protein RegimeFormer subset supplies high-resolution training and residue-scale inference, not the definition of atlas size. Across this hierarchy, perturbation regimes organize damaging, adaptable and uncertain residues, survive matched confounder tests, transfer across protein-language-model backbones, concentrate at independent functional annotations and improve DMS reconstruction. RegimeAtlas therefore extends protein-space organization from sequence families and structural clusters to a global coordinate system for perturbational behaviour.

Perturbation regime is related to, but operates at a different level from, previously described forms of mutational context dependence. Stability-mediated robustness explains why a protein with additional energetic margin can tolerate destabilizing substitutions,\supercite{28,29} and global epistasis describes systematic changes in mutation effect as the genetic background moves across a fitness landscape.\supercite{25,26,27} Direct-coupling and higher-order sequence models likewise capture residue interactions within a family.\supercite{17,19,45} The regime coordinate instead summarizes a cross-protein state that is inferred across heterogeneous proteins and then tested by the organization of their residue-level landscapes. The strong enrichment of fragile residues at active, binding and interface sites, together with the independent alignment with pLDDT and evolutionary conservation, suggests that this state integrates multiple sources of molecular constraint rather than reproducing a single structural or evolutionary descriptor.

The predictive analyses show why this distinction matters. Explicit regime conditioning yielded a modest but reproducible gain on standard ProteinGym activity assays and a larger relative advantage under unseen-protein, unseen-family and low-homology splits. The same coordinate predicted where established models incur larger errors, and uncertainty ranking identified subsets in which performance approached the empirical noise ceiling. These properties support a practical use case in experimental design: the atlas prioritizes residues and mutations for focused measurement across otherwise intractable substitution spaces. The retrospective budget replay and blinded utility analysis both show that this prioritization can recover more beneficial variants at fixed screening cost. This use is complementary to generative protein design and machine-learning-guided directed evolution, which address the separate problem of proposing sequences.\supercite{46,47,48,49,50}

The scale-up also changes how such a resource can be deployed. Materializing $L\times19$ predictions for hundreds of millions of residues would be wasteful for most queries. Storing residue summaries while retaining on-demand substitution inference preserved nearly perfect agreement with the offline model and reduced the storage footprint by more than two orders of magnitude. This architecture makes it possible to couple a broad atlas with focused high-resolution queries. The result is a layered resource: atlas-scale coordinates for navigation, residue tracks for prioritization, substitution-specific predictions for selected sites, and explicit uncertainty at each level.

The protein-to-cell analyses extend the perturbation atlas into a second prediction scale. Protein-derived features improve transcriptomic modelling across multiple datasets, while the drug-response benchmark delineates a harder regime when chemical identity moves far outside training space. The similarity dependence in PRISM shows where richer chemical structure, target engagement, transport and polypharmacology information becomes decisive. This separation is useful for model design: RegimeFormer supplies a strong protein prior, and the next gain in distant chemical space will come from coupling that prior to equally expressive chemical and cellular representations.

Finally, the Explorer turns the learned atlas into an inspectable research system. Protein-level coordinates can be followed directly to residue tracks, individual substitutions, structure overlays, molecular benchmarks and downstream cellular predictions, with versioned provenance and calibrated uncertainty. This is increasingly important as biological AI systems combine models and databases across scales.\supercite{67,68,69,70} The interface converts a cross-protein organizing variable into a queryable resource for selecting experiments, comparing model confidence and tracing predictions across molecular and cellular levels. More broadly, the results establish perturbational response as an organizing axis of protein space that complements sequence, structure and function.

\clearpage
\printbibliography[title={References}]

@article{1,
  title={Proteingym: Large-scale benchmarks for protein fitness prediction and design},
  author={Notin, Pascal and Kollasch, Aaron and Ritter, Daniel and Van Niekerk, Lood and Paul, Steffanie and Spinner, Han and Rollins, Nathan and Shaw, Ada and Orenbuch, Rose and Weitzman, Ruben and others},
  journal={Advances in neural information processing systems},
  volume={36},
  pages={64331--64379},
  year={2023}
}

@article{2,
  title={Deep mutational scanning: a new style of protein science},
  author={Fowler, Douglas M and Fields, Stanley},
  journal={Nature methods},
  volume={11},
  number={8},
  pages={801--807},
  year={2014},
  publisher={Nature Publishing Group US New York}
}

@article{3,
  title={MaveDB: an open-source platform to distribute and interpret data from multiplexed assays of variant effect},
  author={Esposito, Daniel and Weile, Jochen and Shendure, Jay and Starita, Lea M and Papenfuss, Anthony T and Roth, Frederick P and Fowler, Douglas M and Rubin, Alan F},
  journal={Genome biology},
  volume={20},
  number={1},
  pages={223},
  year={2019},
  publisher={Springer}
}

@article{4,
  title={MaveDB 2024: a curated community database with over seven million variant effects from multiplexed functional assays},
  author={Rubin, Alan F and Stone, Jeremy and Bianchi, Aisha Haley and Capodanno, Benjamin J and Da, Estelle Y and Dias, Mafalda and Esposito, Daniel and Frazer, Jonathan and Fu, Yunfan and Grindstaff, Sally B and others},
  journal={Genome biology},
  volume={26},
  number={1},
  pages={13},
  year={2025},
  publisher={Springer}
}

@article{5,
  title={Multiplexed assays of variant effect for clinical variant interpretation},
  author={McEwen, Abbye E and Tejura, Malvika and Fayer, Shawn and Starita, Lea M and Fowler, Douglas M},
  journal={Nature Reviews Genetics},
  volume={27},
  number={2},
  pages={137--154},
  year={2026},
  publisher={Nature Publishing Group UK London}
}

@article{6,
  title={Multiplex assessment of protein variant abundance by massively parallel sequencing},
  author={Matreyek, Kenneth A and Starita, Lea M and Stephany, Jason J and Martin, Beth and Chiasson, Melissa A and Gray, Vanessa E and Kircher, Martin and Khechaduri, Arineh and Dines, Jennifer N and Hause, Ronald J and others},
  journal={Nature genetics},
  volume={50},
  number={6},
  pages={874--882},
  year={2018},
  publisher={Nature Publishing Group US New York}
}

@article{7,
  title={Accurate classification of BRCA1 variants with saturation genome editing},
  author={Findlay, Gregory M and Daza, Riza M and Martin, Beth and Zhang, Melissa D and Leith, Anh P and Gasperini, Molly and Janizek, Joseph D and Huang, Xingfan and Starita, Lea M and Shendure, Jay},
  journal={Nature},
  volume={562},
  number={7726},
  pages={217--222},
  year={2018},
  publisher={Nature Publishing Group UK London}
}

@article{8,
  title={Deep mutational scanning of an RRM domain of the Saccharomyces cerevisiae poly (A)-binding protein},
  author={Melamed, Daniel and Young, David L and Gamble, Caitlin E and Miller, Christina R and Fields, Stanley},
  journal={Rna},
  volume={19},
  number={11},
  pages={1537--1551},
  year={2013},
  publisher={Cold Spring Harbor Laboratory Press}
}

@article{9,
  title={Biological structure and function emerge from scaling unsupervised learning to 250 million protein sequences},
  author={Rives, Alexander and Meier, Joshua and Sercu, Tom and Goyal, Siddharth and Lin, Zeming and Liu, Jason and Guo, Demi and Ott, Myle and Zitnick, C Lawrence and Ma, Jerry and others},
  journal={Proceedings of the national academy of sciences},
  volume={118},
  number={15},
  pages={e2016239118},
  year={2021},
  publisher={National Academy of Sciences}
}

@article{10,
  title={Language models enable zero-shot prediction of the effects of mutations on protein function},
  author={Meier, Joshua and Rao, Roshan and Verkuil, Robert and Liu, Jason and Sercu, Tom and Rives, Alex},
  journal={Advances in neural information processing systems},
  volume={34},
  pages={29287--29303},
  year={2021}
}

@article{11,
  title={Evolutionary-scale prediction of atomic-level protein structure with a language model},
  author={Lin, Zeming and Akin, Halil and Rao, Roshan and Hie, Brian and Zhu, Zhongkai and Lu, Wenting and Smetanin, Nikita and Verkuil, Robert and Kabeli, Ori and Shmueli, Yaniv and others},
  journal={Science},
  volume={379},
  number={6637},
  pages={1123--1130},
  year={2023},
  publisher={American Association for the Advancement of Science}
}

@article{12,
  title={ProtTrans: towards cracking the language of life’s code through self-supervised deep learning and high performance},
  author={Elnaggar, Ahmed and Heinzinger, Michael and Dallago, Christian and others},
  journal={IEEE Trans Pattern analysis and Machine Intelligence},
  volume={14},
  pages={1--16},
  year={2021}
}

@inproceedings{13,
  title={MSA transformer},
  author={Rao, Roshan M and Liu, Jason and Verkuil, Robert and Meier, Joshua and Canny, John and Abbeel, Pieter and Sercu, Tom and Rives, Alexander},
  booktitle={International conference on machine learning},
  pages={8844--8856},
  year={2021},
  organization={PMLR}
}

@inproceedings{15,
  title={Saprot: Protein language modeling with structure-aware vocabulary},
  author={Su, Jin and Han, Chenchen and Zhou, Yuyang and Shan, Junjie and Zhou, Xibin and Yuan, Fajie},
  booktitle={International Conference on Learning Representations},
  volume={2024},
  pages={6987--7009},
  year={2024}
}

@article{17,
  title={Deep generative models of genetic variation capture the effects of mutations},
  author={Riesselman, Adam J and Ingraham, John B and Marks, Debora S},
  journal={Nature methods},
  volume={15},
  number={10},
  pages={816--822},
  year={2018},
  publisher={Nature Publishing Group US New York}
}

@article{18,
  title={Disease variant prediction with deep generative models of evolutionary data},
  author={Frazer, Jonathan and Notin, Pascal and Dias, Mafalda and Gomez, Aidan and Min, Joseph K and Brock, Kelly and Gal, Yarin and Marks, Debora S},
  journal={Nature},
  volume={599},
  number={7883},
  pages={91--95},
  year={2021},
  publisher={Nature Publishing Group UK London}
}

@article{19,
  title={GEMME: a simple and fast global epistatic model predicting mutational effects},
  author={Laine, Elodie and Karami, Yasaman and Carbone, Alessandra},
  journal={Molecular biology and evolution},
  volume={36},
  number={11},
  pages={2604--2619},
  year={2019},
  publisher={Oxford University Press}
}

@article{20,
  title={Accurate proteome-wide missense variant effect prediction with AlphaMissense},
  author={Cheng, Jun and Novati, Guido and Pan, Joshua and Bycroft, Clare and {\v{Z}}emgulyt{\.e}, Akvil{\.e} and Applebaum, Taylor and Pritzel, Alexander and Wong, Lai Hong and Zielinski, Michal and Sargeant, Tobias and others},
  journal={Science},
  volume={381},
  number={6664},
  pages={eadg7492},
  year={2023},
  publisher={American Association for the Advancement of Science}
}

@article{21,
  title={Genome-wide prediction of disease variant effects with a deep protein language model},
  author={Brandes, Nadav and Goldman, Grant and Wang, Charlotte H and Ye, Chun Jimmie and Ntranos, Vasilis},
  journal={Nature genetics},
  volume={55},
  number={9},
  pages={1512--1522},
  year={2023},
  publisher={Nature Publishing Group US New York}
}

@article{22,
  title={Compressing the collective knowledge of ESM into a single protein language model},
  author={Dinh, Tuan and Jang, Seon-Kyeong and Zaitlen, Noah and Ntranos, Vasilis},
  journal={Nature Methods},
  pages={1--13},
  year={2026},
  publisher={Nature Publishing Group US New York}
}

@article{23,
  title={Mega-scale experimental analysis of protein folding stability in biology and design},
  author={Tsuboyama, Kotaro and Dauparas, Justas and Chen, Jonathan and Laine, Elodie and Mohseni Behbahani, Yasser and Weinstein, Jonathan J and Mangan, Niall M and Ovchinnikov, Sergey and Rocklin, Gabriel J},
  journal={Nature},
  volume={620},
  number={7973},
  pages={434--444},
  year={2023},
  publisher={Nature Publishing Group UK London}
}

@article{24,
  title={The genetic architecture of protein stability},
  author={Faure, Andre J and Mart{\'\i}-Aranda, Aina and Hidalgo-Carcedo, Cristina and Beltran, Antoni and Schmiedel, J{\"o}rn M and Lehner, Ben},
  journal={Nature},
  volume={634},
  number={8035},
  pages={995--1003},
  year={2024},
  publisher={Nature Publishing Group UK London}
}

@article{25,
  title={Robustness--epistasis link shapes the fitness landscape of a randomly drifting protein},
  author={Bershtein, Shimon and Segal, Michal and Bekerman, Roy and Tokuriki, Nobuhiko and Tawfik, Dan S},
  journal={Nature},
  volume={444},
  number={7121},
  pages={929--932},
  year={2006},
  publisher={Nature Publishing Group UK London}
}

@article{26,
  title={Global epistasis on fitness landscapes},
  author={Diaz-Colunga, Juan and Skwara, Abigail and Gowda, Karna and Diaz-Uriarte, Ramon and Tikhonov, Mikhail and Bajic, Djordje and Sanchez, Alvaro},
  journal={Philosophical Transactions of the Royal Society B: Biological Sciences},
  volume={378},
  number={1877},
  pages={20220053},
  year={2023}
}

@article{27,
  title={Epistasis in protein evolution},
  author={Starr, Tyler N and Thornton, Joseph W},
  journal={Protein science},
  volume={25},
  number={7},
  pages={1204--1218},
  year={2016},
  publisher={Wiley Online Library}
}

@article{28,
  title={Protein stability promotes evolvability},
  author={Bloom, Jesse D and Labthavikul, Sy T and Otey, Christopher R and Arnold, Frances H},
  journal={Proceedings of the National Academy of Sciences},
  volume={103},
  number={15},
  pages={5869--5874},
  year={2006},
  publisher={National Academy of Sciences}
}

@article{29,
  title={Thermodynamic prediction of protein neutrality},
  author={Bloom, Jesse D and Silberg, Jonathan J and Wilke, Claus O and Drummond, D Allan and Adami, Christoph and Arnold, Frances H},
  journal={Proceedings of the National Academy of Sciences},
  volume={102},
  number={3},
  pages={606--611},
  year={2005},
  publisher={National Academy of Sciences}
}

@article{30,
  title={Diminishing returns epistasis among beneficial mutations decelerates adaptation},
  author={Chou, Hsin-Hung and Chiu, Hsuan-Chao and Delaney, Nigel F and Segr{\`e}, Daniel and Marx, Christopher J},
  journal={Science},
  volume={332},
  number={6034},
  pages={1190--1192},
  year={2011},
  publisher={American Association for the Advancement of Science}
}

@article{31,
  title={Negative epistasis between beneficial mutations in an evolving bacterial population},
  author={Khan, Aisha I and Dinh, Duy M and Schneider, Dominique and Lenski, Richard E and Cooper, Tim F},
  journal={Science},
  volume={332},
  number={6034},
  pages={1193--1196},
  year={2011},
  publisher={American Association for the Advancement of Science}
}

@article{32,
  title={Environmental modulation of global epistasis in a drug resistance fitness landscape},
  author={Diaz-Colunga, Juan and Sanchez, Alvaro and Ogbunugafor, C Brandon},
  journal={Nature communications},
  volume={14},
  number={1},
  pages={8055},
  year={2023},
  publisher={Nature Publishing Group UK London}
}

@article{33,
  title={Highly accurate protein structure prediction with AlphaFold},
  author={Jumper, John and Evans, Richard and Pritzel, Alexander and Green, Tim and Figurnov, Michael and Ronneberger, Olaf and Tunyasuvunakool, Kathryn and Bates, Russ and {\v{Z}}{\'\i}dek, Augustin and Potapenko, Anna and others},
  journal={nature},
  volume={596},
  number={7873},
  pages={583--589},
  year={2021},
  publisher={Nature Publishing Group UK London}
}

@article{34,
  title={AlphaFold Protein Structure Database: massively expanding the structural coverage of protein-sequence space with high-accuracy models},
  author={Varadi, Mihaly and Anyango, Stephen and Deshpande, Mandar and Nair, Sreenath and Natassia, Cindy and Yordanova, Galabina and Yuan, David and Stroe, Oana and Wood, Gemma and Laydon, Agata and others},
  journal={Nucleic acids research},
  volume={50},
  number={D1},
  pages={D439--D444},
  year={2022},
  publisher={Oxford University Press}
}

@article{35,
  title={Fast and accurate protein structure search with Foldseek},
  author={Van Kempen, Michel and Kim, Stephanie S and Tumescheit, Charlotte and Mirdita, Milot and Lee, Jeongjae and Gilchrist, Cameron LM and S{\"o}ding, Johannes and Steinegger, Martin},
  journal={Nature biotechnology},
  volume={42},
  number={2},
  pages={243--246},
  year={2024},
  publisher={Nature Publishing Group US New York}
}

@article{36,
  title={Clustering predicted structures at the scale of the known protein universe},
  author={Barrio-Hernandez, Inigo and Yeo, Jingi and J{\"a}nes, J{\"u}rgen and Mirdita, Milot and Gilchrist, Cameron LM and Wein, Tanita and Varadi, Mihaly and Velankar, Sameer and Beltrao, Pedro and Steinegger, Martin},
  journal={Nature},
  volume={622},
  number={7983},
  pages={637--645},
  year={2023},
  publisher={Nature Publishing Group UK London}
}

@article{37,
  title={UniProt: the universal protein knowledgebase in 2023},
  journal={Nucleic acids research},
  volume={51},
  number={D1},
  pages={D523--D531},
  year={2023},
  publisher={Oxford University Press}
}

@article{38,
  title={Pfam: The protein families database in 2021},
  author={Mistry, Jaina and Chuguransky, Sara and Williams, Lowri and Qureshi, Matloob and Salazar, Gustavo A and Sonnhammer, Erik LL and Tosatto, Silvio CE and Paladin, Lisanna and Raj, Shriya and Richardson, Lorna J and others},
  journal={Nucleic acids research},
  volume={49},
  number={D1},
  pages={D412--D419},
  year={2021},
  publisher={Oxford University Press}
}

@article{39,
  title={InterPro in 2022},
  author={Paysan-Lafosse, Typhaine and Blum, Matthias and Chuguransky, Sara and Grego, Tiago and Pinto, Beatriz L{\'a}zaro and Salazar, Gustavo A and Bileschi, Maxwell L and Bork, Peer and Bridge, Alan and Colwell, Lucy and others},
  journal={Nucleic acids research},
  volume={51},
  number={D1},
  pages={D418--D427},
  year={2023},
  publisher={Oxford University Press}
}

@article{40,
  title={MMseqs2 enables sensitive protein sequence searching for the analysis of massive data sets},
  author={Steinegger, Martin and S{\"o}ding, Johannes},
  journal={Nature biotechnology},
  volume={35},
  number={11},
  pages={1026--1028},
  year={2017},
  publisher={Nature Publishing Group US New York}
}

@misc{41,
  title={Clustering huge protein sequence sets in linear time. Nat Commun 9: 2542},
  author={Steinegger, M and S{\"o}ding, J},
  year={2018}
}

@article{42,
  title={The mutational constraint spectrum quantified from variation in 141,456 humans},
  author={Karczewski, Konrad J and Francioli, Laurent C and Tiao, Grace and Cummings, Beryl B and Alf{\"o}ldi, Jessica and Wang, Qingbo and Collins, Ryan L and Laricchia, Kristen M and Ganna, Andrea and Birnbaum, Daniel P and others},
  journal={Nature},
  volume={581},
  number={7809},
  pages={434--443},
  year={2020},
  publisher={Nature Publishing Group UK London}
}

@article{43,
  title={A high-resolution map of human evolutionary constraint using 29 mammals},
  author={Lindblad-Toh, Kerstin and Garber, Manuel and Zuk, Or and Lin, Michael F and Parker, Brian J and Washietl, Stefan and Kheradpour, Pouya and Ernst, Jason and Jordan, Gregory and Mauceli, Evan and others},
  journal={Nature},
  volume={478},
  number={7370},
  pages={476--482},
  year={2011},
  publisher={Nature Publishing Group UK London}
}

@article{44,
  title={A comparative genomics multitool for scientific discovery and conservation},
  journal={Nature},
  volume={587},
  number={7833},
  pages={240--245},
  year={2020},
  publisher={Nature Publishing Group UK London}
}

@article{45,
  title={Direct-coupling analysis of residue coevolution captures native contacts across many protein families},
  author={Morcos, Faruck and Pagnani, Andrea and Lunt, Bryan and Bertolino, Arianna and Marks, Debora S and Sander, Chris and Zecchina, Riccardo and Onuchic, Jos{\'e} N and Hwa, Terence and Weigt, Martin},
  journal={Proceedings of the National Academy of Sciences},
  volume={108},
  number={49},
  pages={E1293--E1301},
  year={2011},
  publisher={National Academy of Sciences}
}

@article{46,
  title={TadA-Bench: A Million-Variant Benchmark for Future-Round Discovery Toward Agentic Protein Engineering},
  author={Gao, Jin and Zhao, Juntu and Zeng, Zirui and Shen, Jiaqi and Shi, Junhao and Zhao, Dukun and Lu, Yuming and Wang, Dequan},
  journal={arXiv preprint arXiv:2606.02624},
  year={2026}
}

@article{47,
  title={Machine-learning-guided directed evolution for protein engineering},
  author={Yang, Kevin K and Wu, Zachary and Arnold, Frances H},
  journal={Nature methods},
  volume={16},
  number={8},
  pages={687--694},
  year={2019},
  publisher={Nature Publishing Group US New York}
}

@article{48,
  title={Low-N protein engineering with data-efficient deep learning},
  author={Biswas, Surojit and Khimulya, Grigory and Alley, Ethan C and Esvelt, Kevin M and Church, George M},
  journal={Nature methods},
  volume={18},
  number={4},
  pages={389--396},
  year={2021},
  publisher={Nature Publishing Group US New York}
}

@article{49,
  title={Machine learning-assisted directed protein evolution with combinatorial libraries},
  author={Wu, Zachary and Kan, SB Jennifer and Lewis, Russell D and Wittmann, Bruce J and Arnold, Frances H},
  journal={Proceedings of the National Academy of Sciences},
  volume={116},
  number={18},
  pages={8852--8858},
  year={2019},
  publisher={National Academy of Sciences}
}

@article{50,
  title={De novo design of protein structure and function with RFdiffusion},
  author={Watson, Joseph L and Juergens, David and Bennett, Nathaniel R and Trippe, Brian L and Yim, Jason and Eisenach, Helen E and Ahern, Woody and Borst, Andrew J and Ragotte, Robert J and Milles, Lukas F and others},
  journal={Nature},
  volume={620},
  number={7976},
  pages={1089--1100},
  year={2023},
  publisher={Nature Publishing Group UK London}
}

@article{51,
  title={Perturb-Seq: dissecting molecular circuits with scalable single-cell RNA profiling of pooled genetic screens},
  author={Dixit, Atray and Parnas, Oren and Li, Biyu and Chen, Jenny and Fulco, Charles P and Jerby-Arnon, Livnat and Marjanovic, Nemanja D and Dionne, Danielle and Burks, Tyler and Raychowdhury, Raktima and others},
  journal={cell},
  volume={167},
  number={7},
  pages={1853--1866},
  year={2016},
  publisher={Elsevier}
}

@article{53,
  title={Exploring genetic interaction manifolds constructed from rich single-cell phenotypes},
  author={Norman, Thomas M and Horlbeck, Max A and Replogle, Joseph M and Ge, Alex Y and Xu, Albert and Jost, Marco and Gilbert, Luke A and Weissman, Jonathan S},
  journal={Science},
  volume={365},
  number={6455},
  pages={786--793},
  year={2019},
  publisher={American Association for the Advancement of Science}
}

@article{54,
  title={Massively multiplex chemical transcriptomics at single-cell resolution},
  author={Srivatsan, Sanjay R and McFaline-Figueroa, Jos{\'e} L and Ramani, Vijay and Saunders, Lauren and Cao, Junyue and Packer, Jonathan and Pliner, Hannah A and Jackson, Dana L and Daza, Riza M and Christiansen, Lena and others},
  journal={Science},
  volume={367},
  number={6473},
  pages={45--51},
  year={2020},
  publisher={American Association for the Advancement of Science}
}

@article{55,
  title={scPerturb: harmonized single-cell perturbation data},
  author={Peidli, Stefan and Green, Tessa D and Shen, Ciyue and Gross, Torsten and Min, Joseph and Garda, Samuele and Yuan, Bo and Schumacher, Linus J and Taylor-King, Jake P and Marks, Debora S and others},
  journal={Nature Methods},
  volume={21},
  number={3},
  pages={531--540},
  year={2024},
  publisher={Nature Publishing Group US New York}
}

@article{56,
  title={scGen predicts single-cell perturbation responses},
  author={Lotfollahi, Mohammad and Wolf, F Alexander and Theis, Fabian J},
  journal={Nature methods},
  volume={16},
  number={8},
  pages={715--721},
  year={2019},
  publisher={Nature Publishing Group US New York}
}

@article{57,
  title={Predicting transcriptional outcomes of novel multigene perturbations with GEARS},
  author={Roohani, Yusuf and Huang, Kexin and Leskovec, Jure},
  journal={Nature biotechnology},
  volume={42},
  number={6},
  pages={927--935},
  year={2024},
  publisher={Nature Publishing Group US New York}
}

@article{58,
  title={scGPT: toward building a foundation model for single-cell multi-omics using generative AI},
  author={Cui, Haotian and Wang, Chloe and Maan, Hassaan and Pang, Kuan and Luo, Fengning and Duan, Nan and Wang, Bo},
  journal={Nature methods},
  volume={21},
  number={8},
  pages={1470--1480},
  year={2024},
  publisher={Nature Publishing Group US New York}
}

@article{59,
  title={Systema: a framework for evaluating genetic perturbation response prediction beyond systematic variation},
  author={Vi{\~n}as Torn{\'e}, Ramon and Wiatrak, Maciej and Piran, Zoe and Fan, Shuyang and Jiang, Liangze and Teichmann, Sarah A and Nitzan, Mor and Brbi{\'c}, Maria},
  journal={Nature Biotechnology},
  pages={1--10},
  year={2025},
  publisher={Nature Publishing Group US New York}
}

@article{60,
  title={Benchmarking algorithms for generalizable single-cell perturbation response prediction},
  author={Wei, Zhiting and Wang, Yiheng and Gao, Yicheng and Wang, Shuguang and Li, Ping and Si, Duanmiao and Gao, Yuli and Wu, Siqi and Li, Danlu and Dong, Kejing and others},
  journal={Nature Methods},
  volume={23},
  number={2},
  pages={451--464},
  year={2026},
  publisher={Nature Publishing Group US New York}
}

@article{62,
  title={Discovering the anticancer potential of non-oncology drugs by systematic viability profiling},
  author={Corsello, Steven M and Nagari, Rohith T and Spangler, Ryan D and Rossen, Jordan and Kocak, Mustafa and Bryan, Jordan G and Humeidi, Ranad and Peck, David and Wu, Xiaoyun and Tang, Andrew A and others},
  journal={Nature cancer},
  volume={1},
  number={2},
  pages={235--248},
  year={2020},
  publisher={Nature Publishing Group US New York}
}

@article{67,
  title={CellVoyager: AI CompBio agent generates new insights by autonomously analyzing biological data},
  author={Alber, Samuel and Chen, Bowen and Sun, Eric and Isakova, Alina and Wilk, Aaron J and Zou, James},
  journal={Nature Methods},
  volume={23},
  number={4},
  pages={749--759},
  year={2026},
  publisher={Nature Publishing Group US New York}
}

@article{68,
  title={A multi-agent system for automating scientific discovery},
  author={Ghareeb, Ali Essam and Chang, Benjamin and Mitchener, Ludovico and Yiu, Angela and Szostkiewicz, Caralyn J and Shved, Dmytro and Gyimesi, Gavin J and Laurent, Jon M and Wright, Samantha M and Razzak, Muhammed T and others},
  journal={Nature},
  pages={1--3},
  year={2026},
  publisher={Nature Publishing Group UK London}
}

@article{69,
  title={Eubiota: Modular Agentic AI for Autonomous Discovery in the Gut Microbiome},
  author={Lu, Pan and Gao, Yifan and Peng, William G and Zhang, Haoxiang and Zhu, Kunlun and Robinson, Elektra K and Xu, Qixin and Kotaka, Masakazu and Zhang, Harrison G and Li, Bingxuan and others},
  journal={bioRxiv},
  pages={2026--02},
  year={2026},
  publisher={Cold Spring Harbor Laboratory}
}

@article{70,
  title={The Gene Ontology resource: enriching a GOld mine},
  journal={Nucleic acids research},
  volume={49},
  number={D1},
  pages={D325--D334},
  year={2021},
  publisher={Oxford University Press}
}

\clearpage
\section*{Methods}

\subsection*{Study design and evidence hierarchy}
The study was designed to test a protein-level organizing variable through a sequence of increasingly independent evidence layers. Atlas construction and regime discovery used large public sequence collections; residue-level analyses tested whether the learned coordinate localized to known functional sites; DMS and molecular benchmarks provided direct experimental supervision; the atlas-scale scan tested computational scalability and numerical reproducibility; AlphaFold and comparative-genomics analyses provided independent structural and evolutionary evidence; and protein-to-cell analyses tested whether the protein representation contributed information beyond protein identity in downstream perturbation settings. Main-text claims were anchored to the six primary figures, Extended Data provided quality-control and robustness analyses, and the separate Supplementary Information file contains procedural notes and Supplementary Tables 3--23. When the main manuscript cites Supplementary items, the citation is written explicitly as ``Supplementary Table X'' or ``Supplementary Note X'' rather than as a cross-file LaTeX reference.

\subsection*{Sequence sources and atlas assembly}
Raw protein records were aggregated from UniProtKB, RefSeq, Ensembl/GENCODE, MGnify, environmental/metagenomic collections and additional public repositories.\supercite{37,38,39} The raw aggregate contained 424,786,370 entries. Records were normalized to a common schema containing sequence, source accession, source release, taxonomic identifier, protein length and available annotations. The source composition and overlap structure are reported in Supplementary Fig. 1 and Supplementary Note 1. Sequence records were retained only if they used the accepted amino-acid alphabet, contained no internal stop symbol after normalization, exceeded the minimum sequence-length threshold of 30 amino acids, and passed ambiguity and low-complexity filters. Exact duplicates were collapsed by sequence hash. Near-duplicates were clustered at at least 90\% sequence identity using a scalable sequence-clustering workflow based on MMseqs2/Linclust-style operations.\supercite{40,41} Taxonomy was reconciled to a single lineage representation and unresolved or conflicting mappings were flagged. The final atlas contained 202,556,313 distinct sequence entries. All retained sequences entered the protein-level representation, regime-indexing and provenance pipeline that defines RegimeAtlas.

\subsection*{Protein-language-model representation}
All 202,556,313 RegimeAtlas sequences were encoded into the atlas-level ESM-family representation and indexed at the protein level.\supercite{9,10,11} Protein-level embeddings were obtained by pooling residue representations after masking padding and special tokens, and these pooled representations support the atlas-wide regime coordinates, taxonomy/family overlays and retrieval layer. Atlas construction, representation indexing, RegimeFormer optimization, large-scale residue inference and downstream validation were carried out on a 32-GPU NVIDIA A100 cluster over an approximately five-month computational campaign. The production high-resolution residue model used the ESM2 t33 650M UR50D backbone, with a maximum sequence window of 1,022 residues in the streaming implementation. Longer sequences were handled by the windowing policy described in Supplementary Note 2 and audited in Supplementary Fig. 4. A stratified one-million-protein subset of the full atlas was used for RegimeFormer optimization and residue-scale scanning; 995,995 proteins passed that pipeline. For visualization, atlas embeddings were reduced to 50 principal components and projected with UMAP using $n_{\mathrm{neighbors}}=100$, minimum distance 0.4 and random seed 42. The displayed UMAP contains 60,000 sampled points for legibility, while the underlying atlas index contains all 202,556,313 protein entries.

\subsection*{Definition of perturbation-regime coordinates}
Regime coordinates were learned as continuous protein-level variables summarizing the balance among perturbational fragility, adaptability and uncertainty. Coordinate construction used the full atlas-derived protein representation together with perturbation-linked training summaries; the fitted coordinate map was then applied across the 202,556,313-entry atlas index. The resulting variables were standardized before downstream analyses. Discretized regime labels were generated for visualization and stratification. The default display uses ordered bins from fragile through intermediate and adaptive states; the continuous coordinate remained the primary quantity for regression and correlation analyses. Alternative definitions included quantile thresholds, atlas-derived thresholds, wild-type-proxy definitions, pLDDT-supported definitions and entropy-supported definitions. Agreement with the continuous reference is shown in Fig. 5f and Supplementary Fig. 2. Cross-backbone replication was performed independently with ESM-2, ESM-1v and ProtT5 representations, with agreement summarized in Extended Data Table 1.\supercite{9,12}

\subsection*{Confounder and negative-control analyses}
We evaluated whether regime organization could be explained by protein length, amino-acid composition, taxonomy, Pfam family or source batch. The primary confounder analysis used 24,918 proteins with complete covariates. We compared PLM-based prediction of continuous or discretized regime coordinates against models using taxonomy, length and Pfam annotations alone.\supercite{38} Negative controls included length-only representations, shuffled PLM embeddings, random-token embeddings, amino-acid composition, species-only and Pfam-only representations. Two statistics were kept distinct. Figure 1d reports the mean absolute per-protein association, $|\rho|$, aggregated across downstream readouts; because the absolute value is taken before aggregation, this statistic has a positive null floor and superiority is assessed by the real-minus-control difference. Supplementary Fig. 2 reports signed correlations under explicit length matching, length-plus-composition matching and within-group label permutations; these nulls can be centred near zero because sign is retained. Species- and Pfam-matched permutation nulls preserved the corresponding grouping structure while permuting regime labels within matched sets. Empirical $P$ values were estimated from 1,000 permutations. Leave-one-source-out analyses repeated coordinate estimation after removing each major sequence source. Coordinate reproducibility was assessed across independent seeds by Spearman correlation, intraclass correlation, median absolute deviation and RMSE. Details are given in Supplementary Note 2 and Supplementary Fig. 2.

\subsection*{Residue-level fragility, adaptability and uncertainty}
For protein $p$, residue $i$ and candidate substitution $a\neq w_i$, \RegimeFormer{} predicts an effect $\hat y_{i,a}$ and predictive uncertainty $\hat\sigma_{i,a}$. Residue-level fragility $F_i$ was derived from the lower or deleterious tail of the substitution-effect distribution, whereas adaptability $A_i$ summarized the beneficial or tolerated tail. The exact aggregation functions and sign conventions are specified in Supplementary Note 3. Residue-level uncertainty was obtained by aggregating substitution-specific predictive standard deviations. These summaries were used to generate sequence tracks, top-$K$ residue lists and normalized position bins. Full substitution-specific outputs remained available through the on-demand query path.

\subsection*{Functional-site annotation and enrichment}
Functional annotations were harmonized from UniProt, InterPro/Pfam and structure-linked records.\supercite{37,38,39} Categories used in the main enrichment analysis were active sites, ligand-binding sites, protein--protein interfaces, disease-associated missense positions, post-translational modification sites and conserved motifs. For each category, we compared the frequency of annotation among the high-fragility set with the matched residue background and calculated odds ratios, 95\% confidence intervals and two-sided enrichment $P$ values. The reference-matched analysis used the 995,995-protein, 407,048,356-residue scan. Continuous associations between regime score and residue sensitivity were evaluated by Pearson and Spearman correlation and by binned trend analysis. Multiple-testing correction was applied within each enrichment family as described in Supplementary Note 4.

\subsection*{RegimeFormer architecture}
For each candidate substitution, the model input was
\[
 x_{i,a}=f(h_i,e_{w_i},e_a,z_{\mathrm{regime}},q_{\mathrm{assay}},s_i),
\]
where $h_i$ is the residue-level PLM representation, $e_{w_i}$ and $e_a$ are learned embeddings of the wild-type and candidate mutant amino acids, $z_{\mathrm{regime}}$ is the protein-level regime vector, $q_{\mathrm{assay}}$ is an assay-context embedding and $s_i$ contains optional structure or evolutionary features. Inputs were projected to a common latent dimension and passed through six regime-aware attention blocks. Assay context was introduced through a FiLM-style adapter, and soft regime routing was implemented through a gating operator. Output heads predicted continuous effect size, predictive standard deviation and probabilities of damaging, tolerated and beneficial classes. The training objective combined regression and ranking terms with the uncertainty objective. Ablations removed each input or head individually while keeping train/validation splits, optimization budget and evaluation metrics fixed. Hyperparameter sensitivity and seed stability are reported in Supplementary Fig. 3 and Supplementary Note 3.

\subsection*{Experimental DMS and molecular benchmark datasets}
ProteinGym provided the primary standardized DMS benchmark and MaveDB supplied additional activity, binding, stability and abundance measurements.\supercite{1,3,4} The expanded molecular benchmark contains 217 substitution assays and evaluates all reproduced methods on harmonized held-out protein/variant partitions wherever method outputs are available. Task-stratified Spearman correlation is reported for activity, binding, expression, fitness and stability, together with the aggregate across all 217 assays. To prevent proteins represented by many assays from dominating the comparison, Supplementary Fig. 5 additionally nests assay-level results by UniProt protein, yielding 87 proteins for the direct RegimeFormer--Kermut paired analysis. For assays with a compatible physical scale, predictions were calibrated in native units using a linear calibration model $y=\alpha+\beta\hat y$ fitted only on the calibration split. Supplementary Table 8 reports a secondary cross-readout completeness analysis on the subset with harmonized readout labels. Representative avGFP, BLAT, PABP and HRAS matrices are shown in Extended Data Fig. 1 as sequence-level illustrations of the same substitution-reconstruction task.

\subsection*{Baselines and matched comparisons}
Baseline families included ESM/PLM likelihood, DeepSequence/EVE, GEMME, MSA Transformer, Tranception, structure-aware predictors and global or position-only regressors where compatible with the task.\supercite{10,13,17,18,19,20,22} Each head-to-head comparison used the same proteins or residue set, train/validation/test split, model-selection budget and metric definition. Predictor-output and residual-error correlations were used to quantify complementarity, and greedy ensemble selection was performed on validation data only. The full benchmark and ensemble analyses are shown in Supplementary Figs. 5 and 6. The S669 developability/stability analysis was evaluated separately and is reported as an external molecular audit rather than pooled with the ProteinGym activity benchmark.

\subsection*{Out-of-distribution and temporal evaluation}
Out-of-distribution splits were defined at the protein, family and low-homology levels. The low-homology set contained proteins below 30\% sequence identity to the training set. Species-held-out evaluation removed all benchmark proteins from a species before testing that species. Cross-laboratory transfer used reciprocal train/calibrate-on-laboratory-A, test-on-laboratory-B and reverse protocols; transfer retention was defined as cross-laboratory Spearman correlation divided by within-laboratory correlation (Extended Data Table 2). The frozen post-cutoff challenge fixed the checkpoint, thresholds and calibration before evaluation on the post-cutoff assay set and compared paired internal and future Spearman correlations. The prespecified non-inferiority margin was $-0.05$ in $\Delta\rho$ (Supplementary Table 11; Extended Data Fig. 2). No model parameter was updated on the future challenge.

\subsection*{Uncertainty and selective prediction}
Predictive uncertainty was trained jointly with the substitution-effect head. Calibration was assessed with expected calibration error, negative log-likelihood and Brier score. Reliability curves were computed by binning predictions according to confidence or predicted standard deviation and comparing nominal with empirical performance. Selective prediction retained only variants below a series of uncertainty thresholds, producing a risk--coverage curve summarized by selective Spearman correlation, calibration error and the ratio of model--experiment agreement to the experimental noise ceiling (Supplementary Table 7). The paired calibration effect of adding the uncertainty head was evaluated separately within stability, binding, activity, expression and clinical-variant groups (Supplementary Fig. 3).

\subsection*{Atlas-scale streaming inference}
Large-scale inference used the frozen ESM2 t33 650M UR50D backbone and \RegimeFormer{} heads on a one-million-protein proxy set. The production residue scan was distributed across the 32-GPU NVIDIA A100 cluster used for the study and generated 204 top-$K$/binned output shards. This computation formed part of the approximately five-month end-to-end campaign described above. After quality control, 995,995 proteins and 407,048,356 residues were summarized. Stored outputs included per-residue fragility, adaptability and uncertainty, top ten residues per metric, ten normalized position bins, domain-level summaries and regime metadata. Full substitution matrices were not stored. On-demand queries re-ran the substitution head for selected residue--amino-acid pairs. Streaming fidelity was evaluated against an offline reference for ten-bin profiles, domain summaries, individual residue scores and regime tokens. Engine-level timing and failure composition were measured in the atlas-scale inference benchmark (Supplementary Fig. 4), whereas deployed endpoint latency and version-pinned replay were audited separately in Supplementary Fig. 10 and Supplementary Table 21.

\subsection*{Retrospective prioritization and expert utility}
Mutation prioritization was evaluated under fixed candidate budgets of 1, 2, 5, 10, 20 and 40\% of the candidate set, and under fixed substitution counts where appropriate. HitRate@budget measured the fraction of experimentally validated beneficial variants recovered among the screened candidates. Regret was defined relative to the beneficial set left unrecovered at each budget. The chronological replay compared \RegimeFormer{}, the strongest matched baseline, PLM likelihood, conservation and random ranking (Supplementary Table 17; Extended Data Fig. 3). A separate blinded expert utility analysis compared a baseline interface containing sequence, annotations and a conventional predictor with a system-assisted interface containing the same information plus regime coordinates, virtual DMS and uncertainty. Experts selected 20 mutations per protein; HitRate@20, Recall@20, BestFound@20 and Regret@20 were analysed with expert and protein as random effects (Supplementary Table 19).

\subsection*{AlphaFold structural validation}
Structural validation used the 2024-08 AlphaFold DB snapshot.\supercite{33,34} Query proteins were canonicalized and deduplicated, then matched first by exact sequence and subsequently by at least 95\% identity with at least 90\% coverage. Of 1,287,904 non-redundant query proteins, 1,124,782 had an AFDB match, 1,043,915 had a valid structure-linked pLDDT record, and 988,600 passed final QA and backend-indexing criteria. We tested correlations between the regime evidence score and protein-level pLDDT as well as component-specific fragility, adaptability and uncertainty scores. Negative controls included pLDDT-matched sets, length-matched sets, within-protein shuffling and a global-mean pLDDT baseline. pLDDT was analysed as AlphaFold structural confidence; thermodynamic stability was treated as a separate molecular phenotype.

\subsection*{Evolutionary constraint analysis}
Evolutionary evidence combined residue-level phyloP conservation with codon-level ortholog analysis.\supercite{42,43,44} Starting from 57,048 phyloP-matched positions, transcripts with complete coding sequence and frame consistency were retained, yielding 51,872 CDS-ready positions. Strict one-to-one ortholog filtering produced 46,118 positions. For the gene-family analysis, 18,624 candidate orthogroups were reduced to 8,014 codon-aligned QC-passing families after single-copy, CDS, strict orthology and frame-quality filters. Median ortholog depth was 18 species with a minimum of 12. We estimated $\omega=d_N/d_S$ using PAML M0, PAML free-ratio, HyPhy FEL, HyPhy BUSTED and NG86 counting, and examined sensitivity to ortholog depth and gene-level bootstrap resampling (Supplementary Fig. 9; Supplementary Note 6).

\subsection*{Protein-to-cell knowledge graph and perturbation datasets}
Three nested data objects were maintained for portal integration and benchmark evaluation. The Fig. 6 benchmark bridge is a restricted induced subgraph containing only entities with complete model inputs and response labels; it contains 467,150 matched bridge records, 4,137 nodes and 269,337 edges, with 477 proteins, 38 drugs, 563 cell lines and 93 genes. Protein identifiers were normalized to UniProt, drugs to canonical chemical identifiers and cell lines to a common ontology where possible. Separately, the portal-wide knowledge graph integrates proteins, drugs, cell types, pathways and tissues across the broader RegimeAtlas resource; its counts and matching rates are reported in Supplementary Fig. 7a,b and Supplementary Note 7 and provide the broader resource layer surrounding the restricted benchmark bridge. The upstream cellular-response corpus contains 3,176,765 profiles across SciPlex3, Norman, Dixit, LINCS and Tahoe; Supplementary Fig. 7c--h reports its dataset and cell-type coverage, genetic/chemical perturbation composition, split manifests and entity-mapping missingness, residual distributions, perturbation-level calibration and uncertainty, and top-50 DEG recovery. Only the subset satisfying the bridge matching and feature-completeness criteria enters the 467,150 benchmark records. Transcriptomic perturbation benchmarks were drawn from SciPlex, Norman and Dixit/Perturb-seq resources, with harmonized processing guided by scPerturb.\supercite{51,53,54,55} Specialized baselines included GEARS, scGPT and related perturbation models where compatible.\supercite{56,57,58}

\subsection*{Transcriptomic perturbation evaluation}
Transcriptomic predictions were evaluated with RMSE, Pearson correlation and top-50 differentially expressed gene Jaccard overlap. Metrics were computed per perturbation and summarized at the dataset level. Basal-only prediction was included as a control. Cell-type residuals were examined separately for A549, K562 and MCF7 in SciPlex3. We also inspected the distribution of top-50 DEG overlap to distinguish broad expression-profile agreement from recovery of perturbation-specific genes. This evaluation follows recent recommendations to avoid relying on a single mean-expression metric for perturbation response prediction.\supercite{59,60}

\subsection*{Drug-response out-of-distribution evaluation}
Drug-response evaluation used PRISM 25Q2 records with AUC and log2-IC50 endpoints.\supercite{62} After dose and viability quality control, 1,101,954 rows across 727 cell lines and 1,482 compounds remained. We constructed random, unseen-cell-line and unseen-compound splits with zero direct overlap in the held-out entity. Compound similarity was measured using Tanimoto similarity of the chemical representation, and performance was stratified into similarity bins above 0.5, 0.3--0.5 and below 0.3. We report $R^2$, Pearson and Spearman correlation, ECE and negative log-likelihood. The unseen-compound split was retained as an explicit distribution-shift analysis rather than pooled with the easier random and unseen-cell settings (Supplementary Fig. 8; Supplementary Note 8).

\subsection*{RegimeAtlas Explorer, versioning and grounding audit}
The Explorer exposes versioned interfaces for protein lookup, regime navigation, virtual DMS, structure visualization, cellular perturbation, spatial ligand--receptor analysis and bulk download. Each output records data version, model checkpoint and provenance metadata. The grounding audit compared an LLM-only system, retrieval-augmented generation and the full RegimeAtlas agent on task accuracy, citation precision and recall, unsupported-claim rate, tool-call success and run-to-run intraclass correlation (Supplementary Table 20). Portal numerical reproducibility was checked by comparing streaming or API outputs with the offline reference and by replaying identical queries under a pinned version (Supplementary Table 21). The core platform views in Extended Data Figs. 5--8 expose this same versioned backend, while Supplementary Figs. 11 and 12 provide the detailed TP53 Assistant and three-dimensional structure views; backend and spatial-resource audits are reported in Supplementary Fig. 10. Agentic workflow design was informed by recent systems that separate planning, execution, verification and evidence-grounded synthesis.\supercite{67,68,69}

\subsection*{Spatial communication demonstration}
The spatial demonstration used the 10x Visium V1 Human Lymph Node dataset and a curated ligand--receptor resource assembled from CellPhoneDB, CellChat, NicheNet, Connectome and LIANA-consensus sources. Forty-eight spatial accessions were screened in the broader resource audit; 39 passed format, reference-alignment and spot-level QC. For the displayed lymph-node case, approximately 4,035 spots and seven cell types were retained. Communication scores were evaluated across a range of minimum-expression thresholds; the median stability ICC across parameter sweeps was 0.90, rank stability was $\tau=0.94$, and 92.6\% of top pairs were retained under the default robust setting (Supplementary Fig. 10).

\subsection*{Statistics and reproducibility}
Unless otherwise stated, reported correlations are Spearman $\rho$ for rank-based analyses and Pearson $r$ for linear association. Confidence intervals for aggregate performance differences were obtained by paired bootstrap resampling at the protein or assay level, matching the unit of independence in the corresponding experiment. Enrichment confidence intervals were calculated on the matched residue sets. Multiple comparisons in the cross-analysis summary were adjusted using the Benjamini--Hochberg procedure (Supplementary Table 22). All random splits and visualization subsamples used fixed seeds recorded in the source data. Five-seed model stability, leave-one-source-out atlas stability, cross-backbone replication, strict OOD splits and definition sensitivity are reported as separate, interpretable robustness analyses.

\subsection*{Data availability}
The study uses the public sequence, structure, DMS, perturbation and drug-response resources cited in the Methods. Source data underlying Figs. 1--6, Extended Data Figs. 1--8, Extended Data Tables 1--2, Supplementary Figs. 1--12 and Supplementary Tables 3--23 are provided with the manuscript. The RegimeAtlas release used for the reported analyses contains the 202,556,313-entry protein-level atlas index and provenance schema, together with residue-level summaries for the 995,995-protein high-resolution scan and the benchmark manifests used throughout the study.

\subsection*{Code availability}
The analysis code covers regime-coordinate construction, RegimeFormer training and evaluation, atlas-scale inference, structural and evolutionary validation, figure generation and portal-query reproduction. The study materials retain the analysis environment specification and checkpoint identifiers used for the reported results.

\section*{Additional information}
Supplementary Information accompanies this manuscript and contains Supplementary Results, Supplementary Figs. 1--12, Supplementary Notes and Supplementary Tables 3--23.

\clearpage
\section*{Extended Data}

Extended Data Figs. 1--4 extend the quantitative validation of RegimeFormer through substitution-matrix reconstruction, frozen temporal transfer, fixed-budget experimental prioritization and held-out predictor-error forecasting. Extended Data Figs. 5--8 document the core versioned RegimeAtlas interfaces spanning atlas provenance, residue-level querying, protein-to-cell prediction and on-demand virtual DMS. Supplementary Figs. 11 and 12 provide the detailed TP53 Assistant and three-dimensional structure-workspace views. Extended Data Tables 1 and 2 report cross-backbone regime replication and cross-laboratory/cross-protocol transfer, respectively.

\clearpage
\subsection*{Extended Data Fig. 1: experimental DMS recovery resolves local landscape structure}
The purpose of this analysis is to test whether the aggregate benchmark in Fig. 3 corresponds to recognizable substitution-level structure within individual proteins. The four examples span fluorescence, antibiotic resistance, RNA binding and GTPase activity, so agreement cannot be attributed to a single assay scale. RegimeFormer reproduces broad residue-specific bands and the relative ordering of many substitution effects, with per-protein Spearman correlations from 0.636 to 0.748 and RMSE from 0.191 to 0.242. The observed-versus-predicted clouds also show residual scatter, which is expected from assay noise and is quantified separately by the experimental noise-ceiling analysis. Thus, Extended Data Fig. 1 provides a local, matrix-level view of the same reconstruction problem summarized across assays in Fig. 3 rather than serving as a second benchmark.

\clearpage
\thispagestyle{plain}

\begin{center}
\makebox[\textwidth][c]{%
\includegraphics[
    width=0.94\textwidth,
    height=0.82\textheight,
    keepaspectratio
]{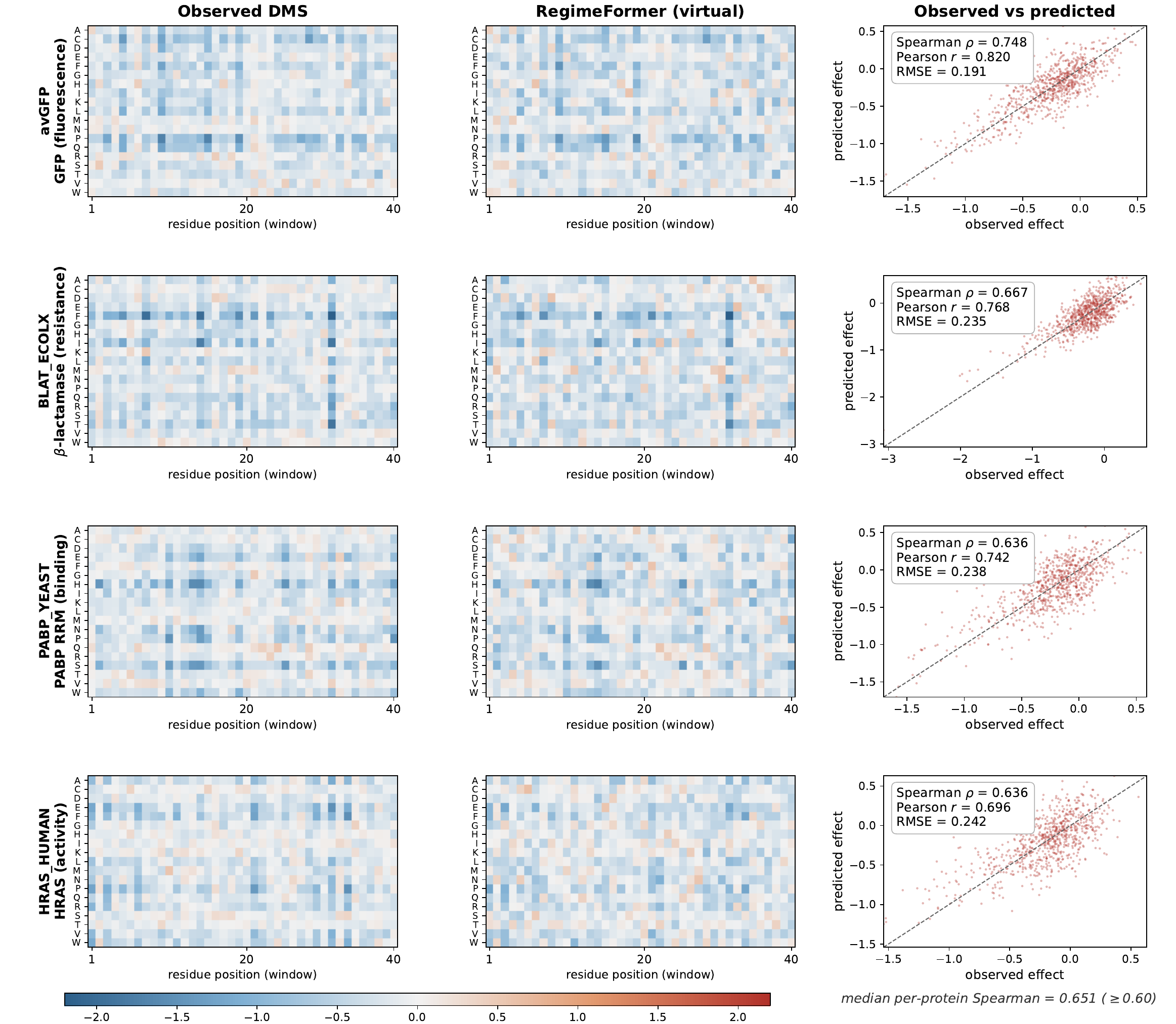}%
}
\end{center}

\vspace{0.3em}

\captionsetup{
    font=small,
    justification=justified,
    singlelinecheck=false
}

\captionof*{figure}{\textbf{Extended Data Fig. 1 | Experimental DMS landscapes are recovered at substitution resolution.} Observed deep-mutational-scanning matrices are compared with RegimeFormer virtual matrices for avGFP fluorescence, BLAT antibiotic resistance, PABP RNA-binding and HRAS activity. The corresponding observed-versus-predicted substitution plots give Spearman correlations of 0.748, 0.667, 0.636 and 0.636, Pearson correlations of 0.820, 0.768, 0.742 and 0.696, and RMSE values of 0.191, 0.235, 0.238 and 0.242, respectively. The representative examples show that the model recovers residue-specific bands and local high-effect regions across distinct assay modalities rather than only reproducing an assay-level average. The 217-assay benchmark in Fig. 3 provides the population-level molecular comparison, while the nested 87-protein analysis in Supplementary Fig. 5 tests whether the same result persists after preventing assay-rich proteins from dominating the aggregate.}

\clearpage
\subsection*{Extended Data Fig. 2: the frozen model retains rank performance on later-release assays}
This experiment addresses temporal robustness rather than random-split generalization. The checkpoint, thresholds and calibration are held fixed and later-release assays are evaluated without updating the model. Most assays lie close to the identity line, and the paired mean change is only $-0.006$ in Spearman correlation, with a 95\% CI of $[-0.020,0.007]$. Because the lower confidence bound remains above the prespecified $-0.05$ non-inferiority margin, the aggregate model performance is retained across the temporal shift. The scatter around the identity line remains informative: individual assays can improve or deteriorate, showing that temporal robustness is a population-level property rather than a guarantee for every assay.

\clearpage
\thispagestyle{plain}

\begin{center}
\makebox[\textwidth][c]{%
\includegraphics[
    width=0.94\textwidth,
    height=0.82\textheight,
    keepaspectratio
]{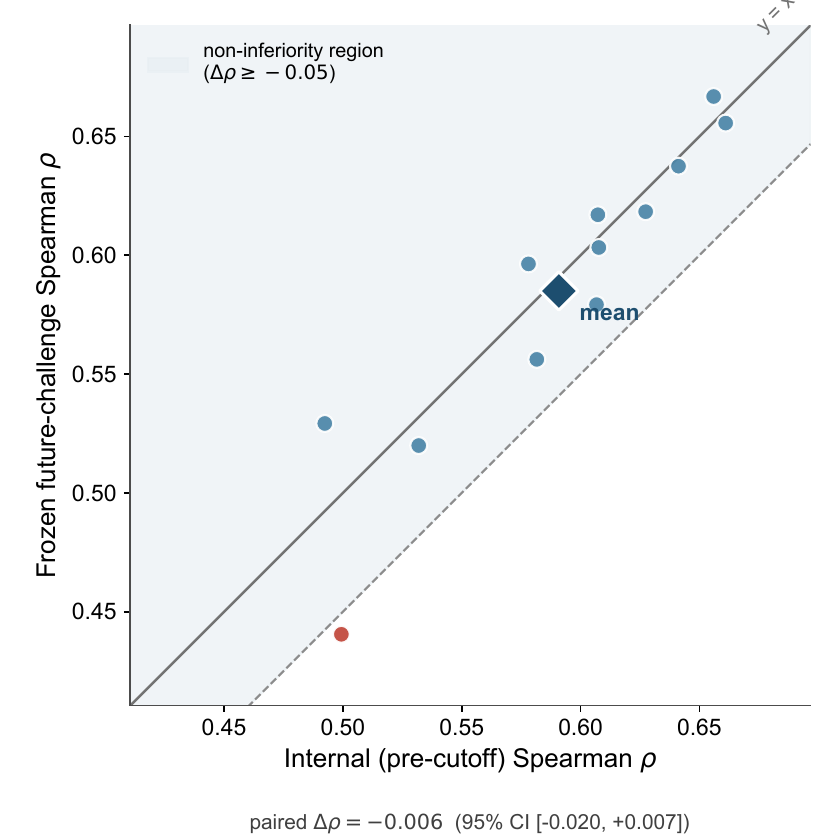}%
}
\end{center}

\vspace{0.3em}

\captionsetup{
    font=small,
    justification=justified,
    singlelinecheck=false
}

\captionof*{figure}{\textbf{Extended Data Fig. 2 | A frozen post-cutoff challenge shows little temporal degradation.} Each point compares internal pre-cutoff Spearman correlation with performance on a later-release assay evaluated using the frozen checkpoint, thresholds and calibration. The paired mean change is $\Delta\rho=-0.006$ (95\% CI, $-0.020$ to 0.007), remaining above the prespecified non-inferiority boundary of $-0.05$. Individual assays vary around the identity line, but the aggregate result indicates that the measured DMS performance is not dependent on re-tuning the model after observing the later-release benchmark.}

\clearpage
\subsection*{Extended Data Fig. 3: the perturbation map is most useful when experiments are scarce}
The fixed-budget replay converts prediction quality into an experimental-selection question. At a 5\% screening budget, RegimeFormer recovers a HitRate of 0.540 compared with 0.434 for the strongest baseline, a 24\% relative gain. The corresponding regret curve falls fastest for the regime-aware ranking at low and intermediate budgets. The convergence of all methods near exhaustive screening is expected: once nearly every candidate is measured, ranking contributes little. The separation at 1--20\% budgets therefore identifies the operational regime in which the model can reduce experimental search rather than merely reproduce a retrospective correlation.

\clearpage
\begin{landscape}
\thispagestyle{plain}

\begin{center}
\makebox[\linewidth][c]{%
\includegraphics[
    width=1.05\linewidth,
    height=0.96\textheight,
    keepaspectratio
]{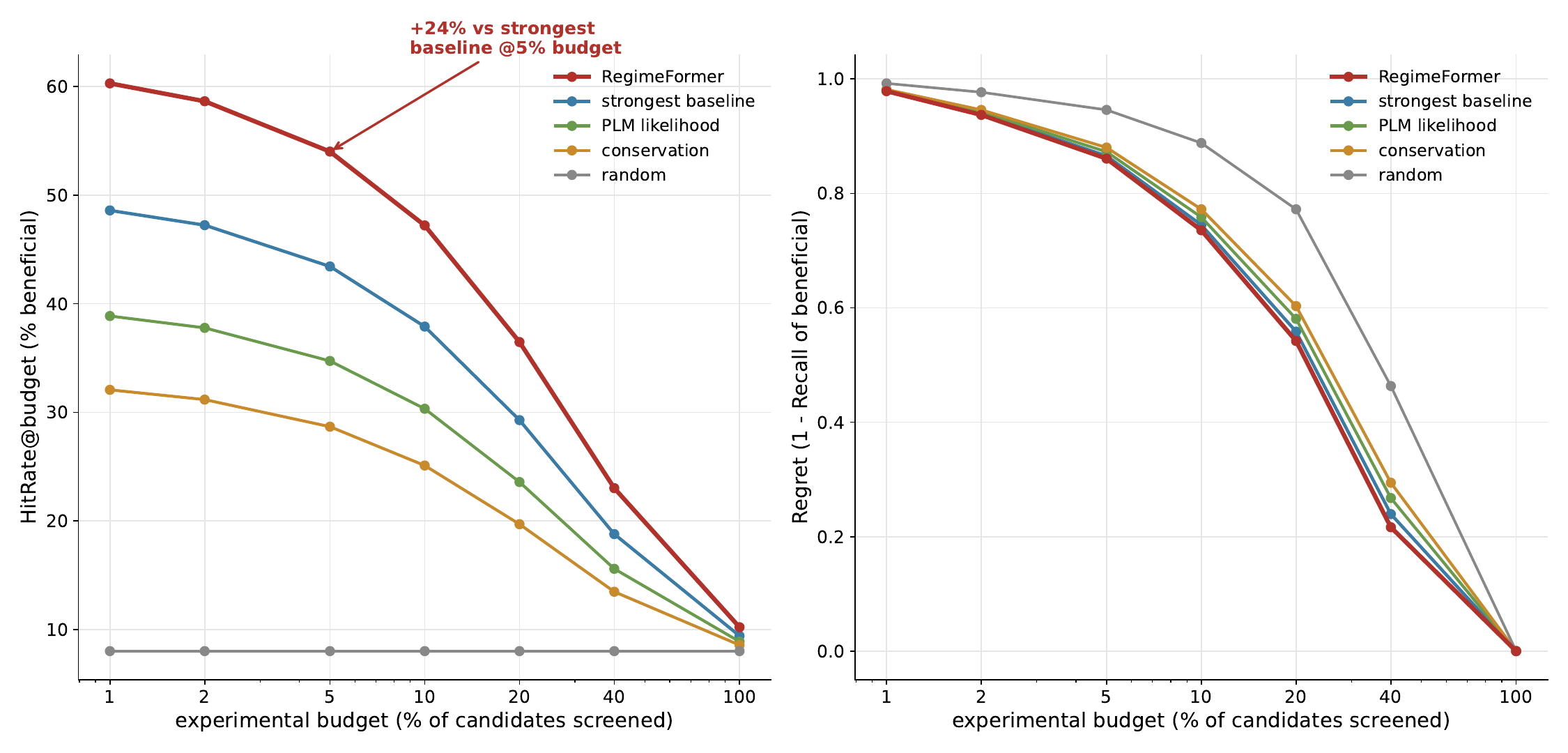}%
}
\end{center}

\vspace{0.3em}

\captionsetup{
    font=small,
    justification=justified,
    singlelinecheck=false
}

\captionof*{figure}{\textbf{Extended Data Fig. 3 | Regime-aware ranking improves mutation recovery under limited experimental budgets.} Left, HitRate among experimentally validated beneficial variants as the screened fraction of candidates increases. At a 5\% budget, RegimeFormer reaches a HitRate of 0.540 versus 0.434 for the strongest baseline, a 24\% relative increase. Right, regret, defined as one minus recall of beneficial variants, falls more rapidly for the regime-aware ranking in the low- and intermediate-budget range. The curves converge as the screen approaches exhaustive coverage, locating the practical advantage in the experimentally relevant setting in which only a small fraction of variants can be tested.}

\end{landscape}
\clearpage
\subsection*{Extended Data Fig. 4: regime state predicts model difficulty beyond ordinary metadata}
Here the target is not the RegimeFormer mutation score itself, but the error made by held-out variant-effect predictor families. Metadata alone yields Spearman 0.310 with observed predictor error, while regime alone yields 0.380. Combining the two increases the held-out correlation to 0.533 and identifies high-error cases with AUROC 0.861. The calibration curve rises monotonically with predicted failure probability, showing that the score is informative not only for ranking errors but also for stratifying high-risk predictions. This result separates two uses of uncertainty: RegimeFormer has its own predictive-uncertainty head, whereas the failure model estimates when an external predictor is likely to be unreliable.

\clearpage
\begin{landscape}
\thispagestyle{plain}

\begin{center}
\makebox[\linewidth][c]{%
\includegraphics[
    width=1.05\linewidth,
    height=0.96\textheight,
    keepaspectratio
]{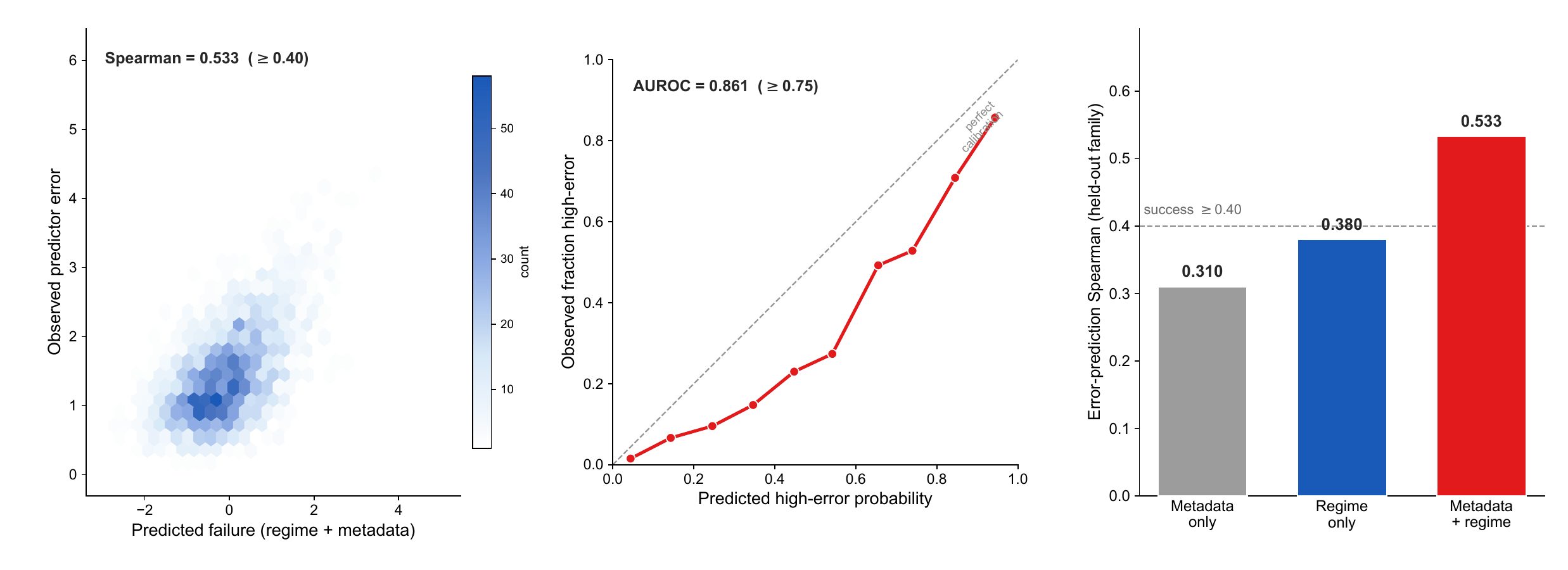}%
}
\end{center}

\vspace{0.3em}

\captionsetup{
    font=small,
    justification=justified,
    singlelinecheck=false
}

\captionof*{figure}{\textbf{Extended Data Fig. 4 | Regime state forecasts where independent variant-effect predictors fail.} Left, a metadata-plus-regime failure score correlates with observed predictor error at Spearman 0.533 on held-out predictor families. Middle, predicted high-error probability increases monotonically with the observed fraction of high-error cases and yields AUROC 0.861. Right, metadata alone reaches error-prediction Spearman 0.310, regime alone 0.380 and the combined model 0.533. Thus, the regime coordinate carries a component of model-difficulty information that is complementary to ordinary assay and protein metadata and can be used as a quality-control prior before new DMS labels are available.}

\end{landscape}
\clearpage
\subsection*{Extended Data Fig. 5: the atlas interface links model access to provenance}
The landing page and methods/provenance view are paired because they represent two sides of the same resource. The first exposes the task-facing entry points---Assistant, Protein Explorer, Regime UMAP, Virtual DMS and API---whereas the second exposes the frozen atlas scale, release version, data sources and model summary. This pairing is important for the paper's large-model framing: RegimeAtlas is not presented as a collection of disconnected web demos, but as an access layer over a versioned RegimeFormer evidence object. The methods view also distinguishes the full atlas from the coordinate-analysis subset and from downstream residue/cellular prediction heads, preventing the interface from conflating these different data scales.

\clearpage
\begin{landscape}
\thispagestyle{plain}

\begin{center}
\textbf{a}\par\vspace{0.2em}
\includegraphics[
    width=0.99\linewidth,
    height=0.42\textheight,
    keepaspectratio
]{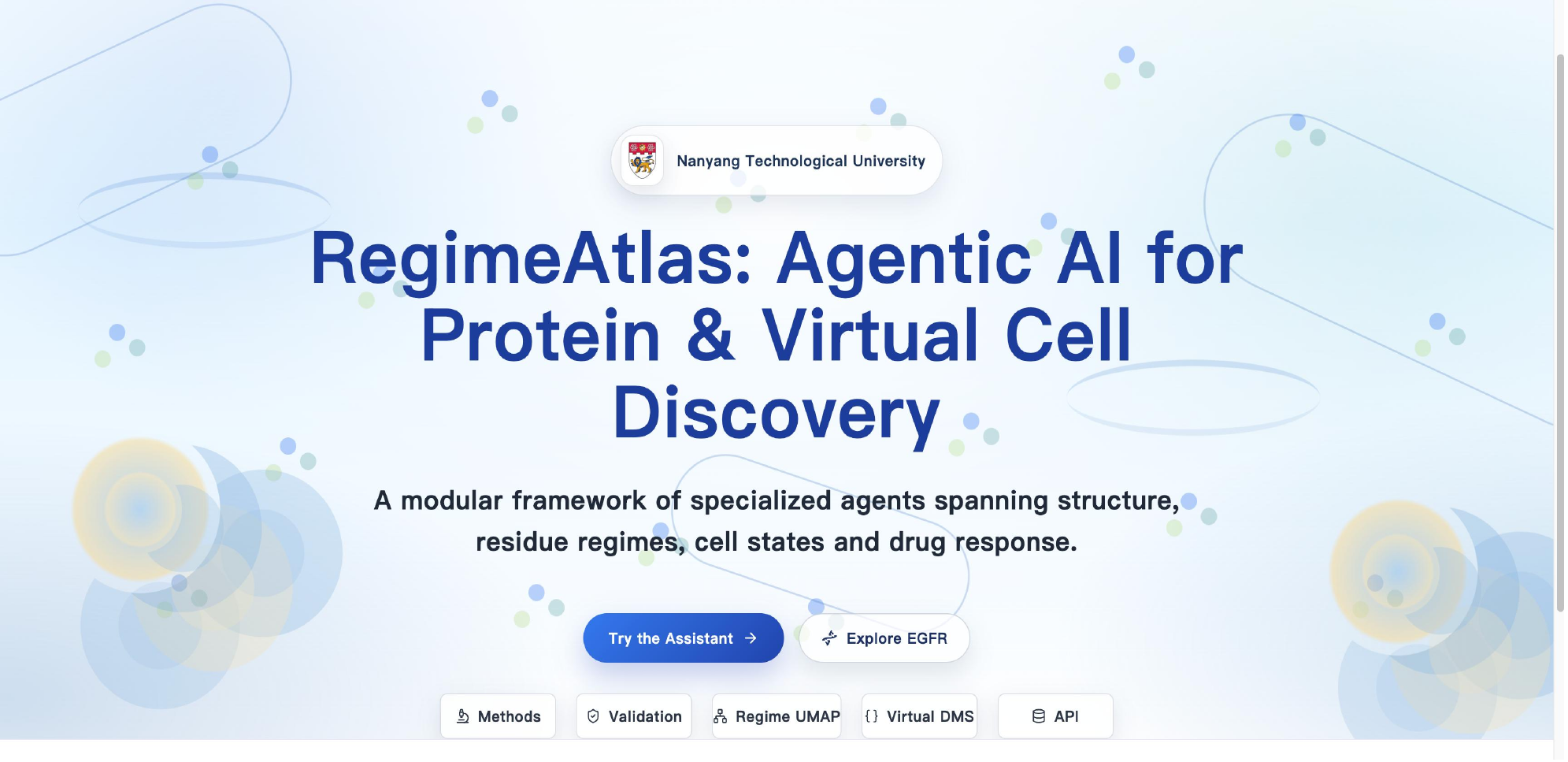}

\vspace{0.6em}

\textbf{b}\par\vspace{0.2em}
\includegraphics[
    width=0.99\linewidth,
    height=0.42\textheight,
    keepaspectratio
]{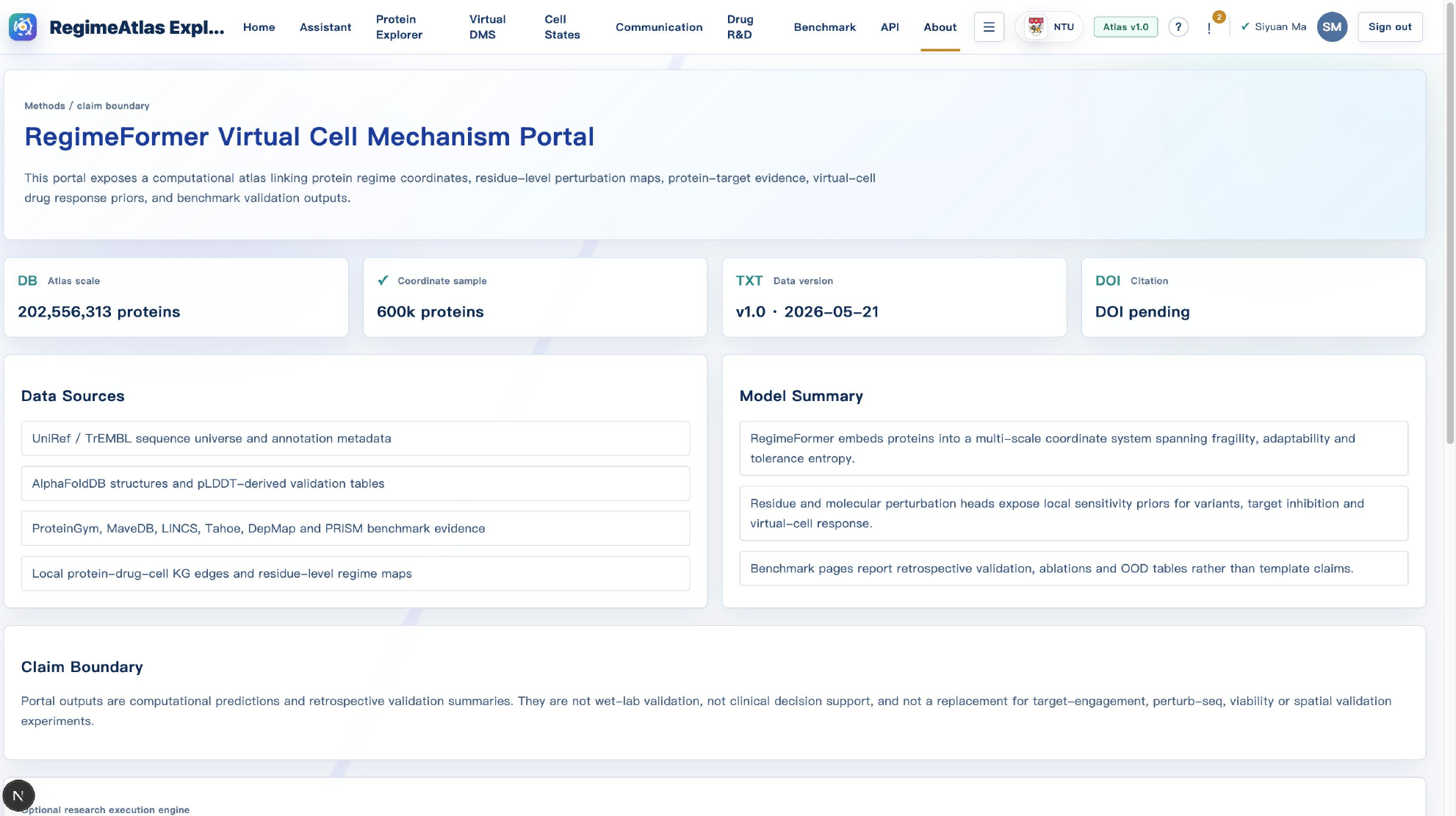}
\end{center}

\vspace{0.3em}

\captionsetup{
    font=small,
    justification=justified,
    singlelinecheck=false
}

\captionof*{figure}{\textbf{Extended Data Fig. 5 | RegimeAtlas entry point, provenance and model scope.} \textbf{a}, The landing page exposes a common entry point to the Assistant, Protein Explorer, Regime UMAP, Virtual DMS and API, organizing the user workflow around a shared RegimeFormer evidence object. \textbf{b}, The methods/provenance page exposes the full 202,556,313-protein atlas, coordinate-analysis subset, release version and contributing data sources, and separates protein-level regime coordinates from residue and cellular prediction heads. Pairing the two views shows how discovery-oriented navigation is coupled to an inspectable data and model provenance layer.}

\end{landscape}
\clearpage
\subsection*{Extended Data Fig. 6: EGFR queries expose a structured residue-ranking object}
The EGFR example illustrates how a natural-language request is translated into explicit model fields rather than answered by free-form generation alone. The returned table contains sequence position, wild-type residue, sensitivity score and the substitutions assigned the largest deleterious effects. This view is therefore the decision-layer projection of the residue-level perturbation map: it reduces a high-dimensional mutation landscape to a ranked set of candidate residues while preserving the quantitative variables needed to reproduce or inspect the ranking. The screenshot is an interface demonstration; its scientific support comes from the DMS, calibration and prioritization analyses elsewhere in the paper.

\clearpage
\begin{landscape}
\thispagestyle{plain}

\begin{center}
\makebox[\linewidth][c]{%
\includegraphics[
    width=1.05\linewidth,
    height=0.96\textheight,
    keepaspectratio
]{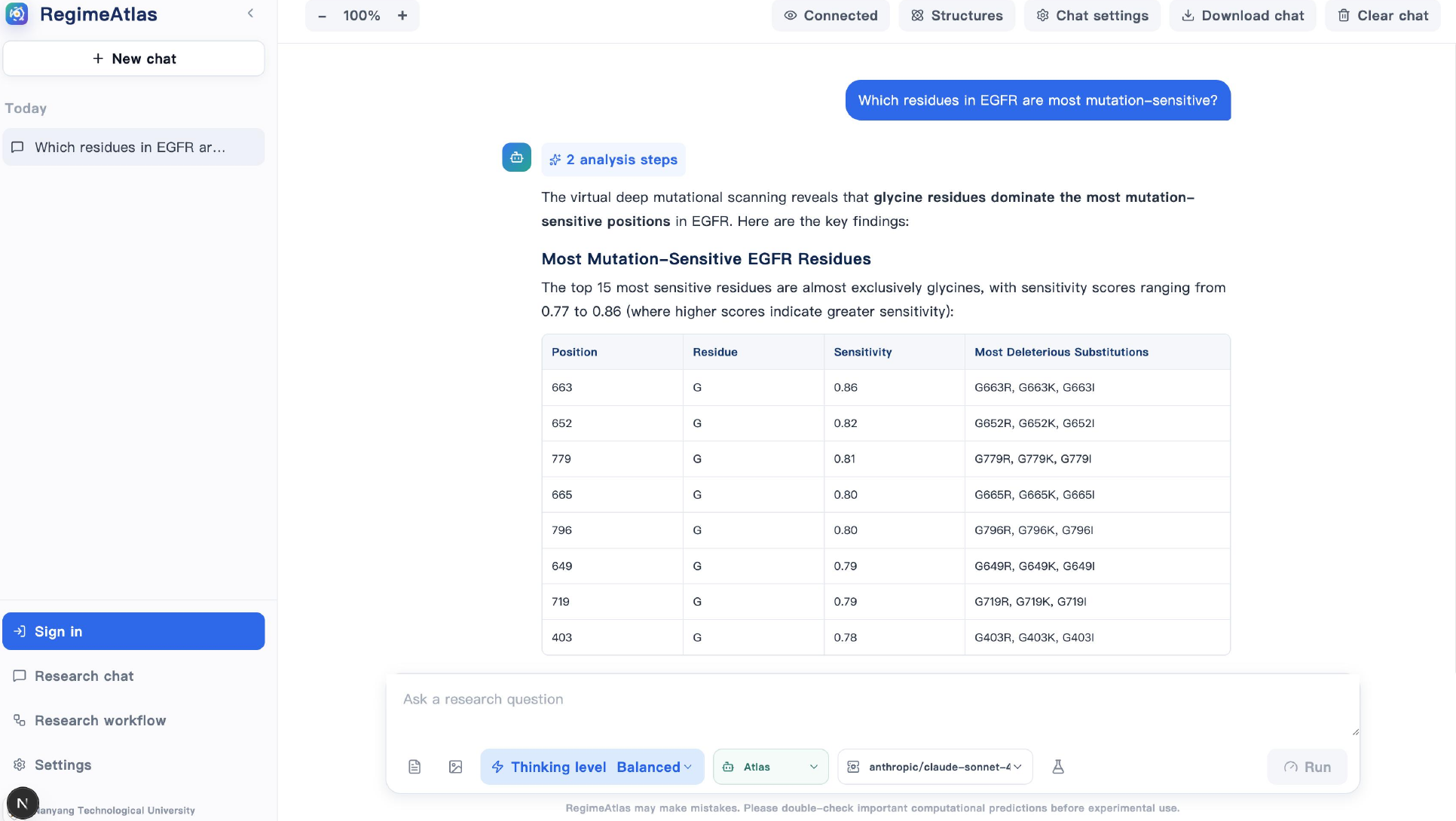}%
}
\end{center}

\vspace{0.3em}

\captionsetup{
    font=small,
    justification=justified,
    singlelinecheck=false
}

\captionof*{figure}{\textbf{Extended Data Fig. 6 | Residue-level RegimeFormer query for EGFR.} A natural-language request is resolved into a ranked mutation-sensitivity table containing residue position, wild-type identity, sensitivity score and the substitutions with the largest predicted deleterious effects. The view converts the protein-level regime representation into a residue-level decision surface while keeping the ranking quantitatively inspectable rather than returning a text-only answer.}

\end{landscape}
\clearpage
\subsection*{Extended Data Fig. 7: the protein-to-cell interface exposes prediction and uncertainty together}
The EGFR-inhibition example shows the final cross-scale query layer. The interface returns a predicted fitness effect of 0.53 with a 95\% interval of 0.27--0.79, together with an uncertainty summary, supporting-evidence count and validation-source classes. The width of the displayed interval is itself informative: the interface is designed to expose model uncertainty rather than convert a cross-scale prediction into a deterministic statement. This view operationalizes Fig. 6 by placing molecular prior, cellular response and evidence provenance in one query object; benchmark support remains the dataset-level transcriptomic and drug-response analyses in Fig. 6 and Supplementary Figs. 7--8.

\clearpage
\begin{landscape}
\thispagestyle{plain}
\begin{center}
\makebox[\linewidth][c]{%
\includegraphics[
    width=1.05\linewidth,
    height=0.96\textheight,
    keepaspectratio
]{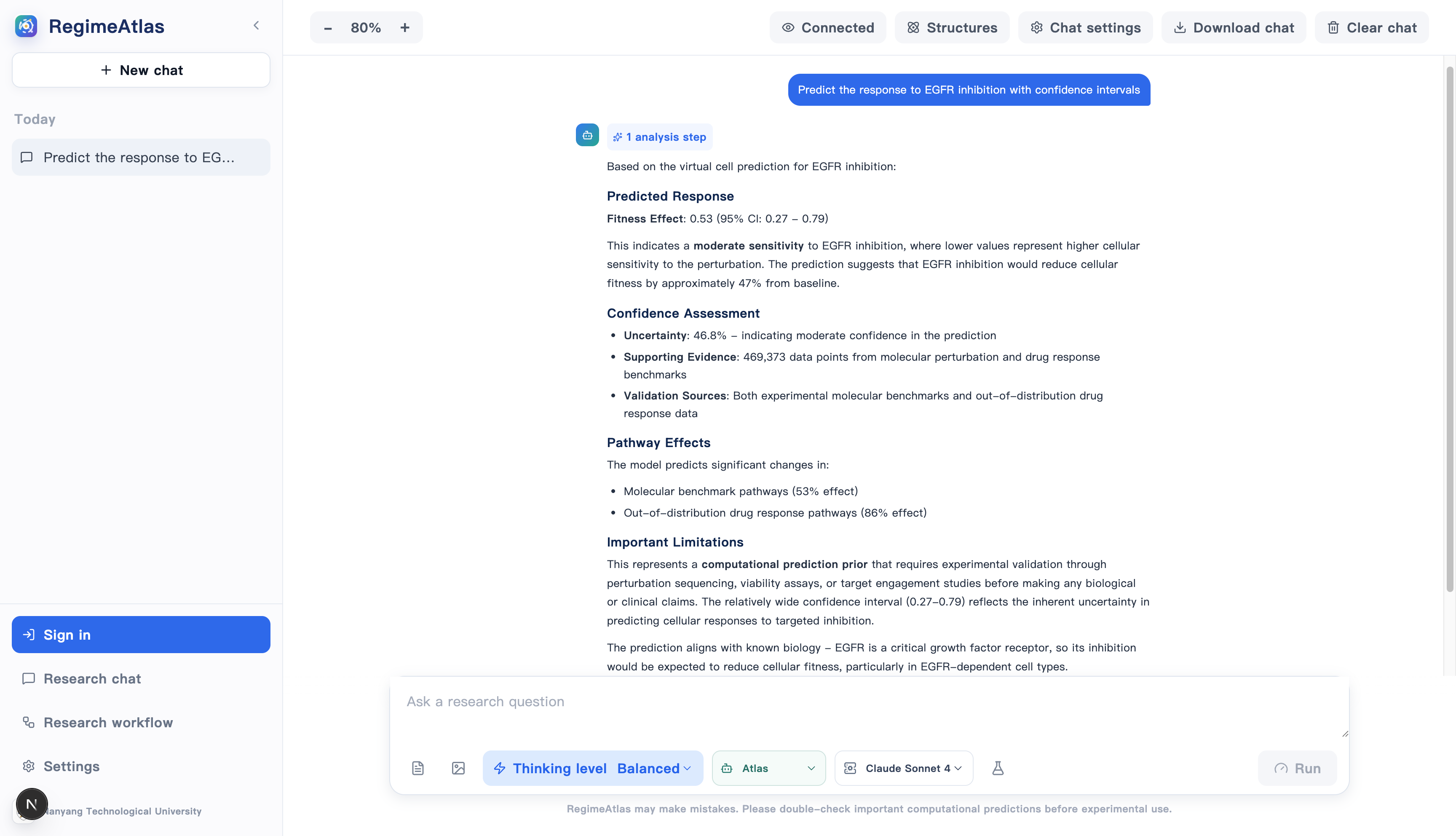}%
}
\end{center}
\end{landscape}
\clearpage
\captionsetup{font=small,justification=justified,singlelinecheck=false}
\captionof*{figure}{\textbf{Extended Data Fig. 7 | Protein-to-cell query with explicit predictive uncertainty.} For an EGFR-inhibition query, the interface reports a predicted fitness effect of 0.53 with a 95\% interval of 0.27--0.79, an uncertainty estimate and the evidence classes contributing to the response. This view exposes the protein-to-cell bridge, uncertainty layer and provenance information together, connecting the model-level analysis in Fig. 6 with a user-facing query rather than treating the displayed prediction as a separate benchmark.}
\clearpage
\subsection*{Extended Data Fig. 8: the Virtual DMS Studio exposes the full substitution matrix on demand}
The DMS Studio is the highest-resolution user-facing view of the protein perturbation model. A UniProt accession or raw sequence is converted into the 20-amino-acid substitution matrix for a selected sequence window, with the protein-level sensitivity summary retained above the matrix. This design mirrors the storage strategy in Fig. 4: atlas-scale residue summaries are materialized broadly, whereas full $L\times19$ substitution predictions are generated only when requested. Extended Data Fig. 8 therefore links the engineering architecture of the atlas to its scientific output, showing how a user can move from a protein identifier to a mutation-resolved landscape without storing every possible matrix in advance.

\clearpage
\begin{landscape}
\thispagestyle{plain}
\begin{center}
\makebox[\linewidth][c]{%
\includegraphics[
    width=1.05\linewidth,
    height=0.96\textheight,
    keepaspectratio
]{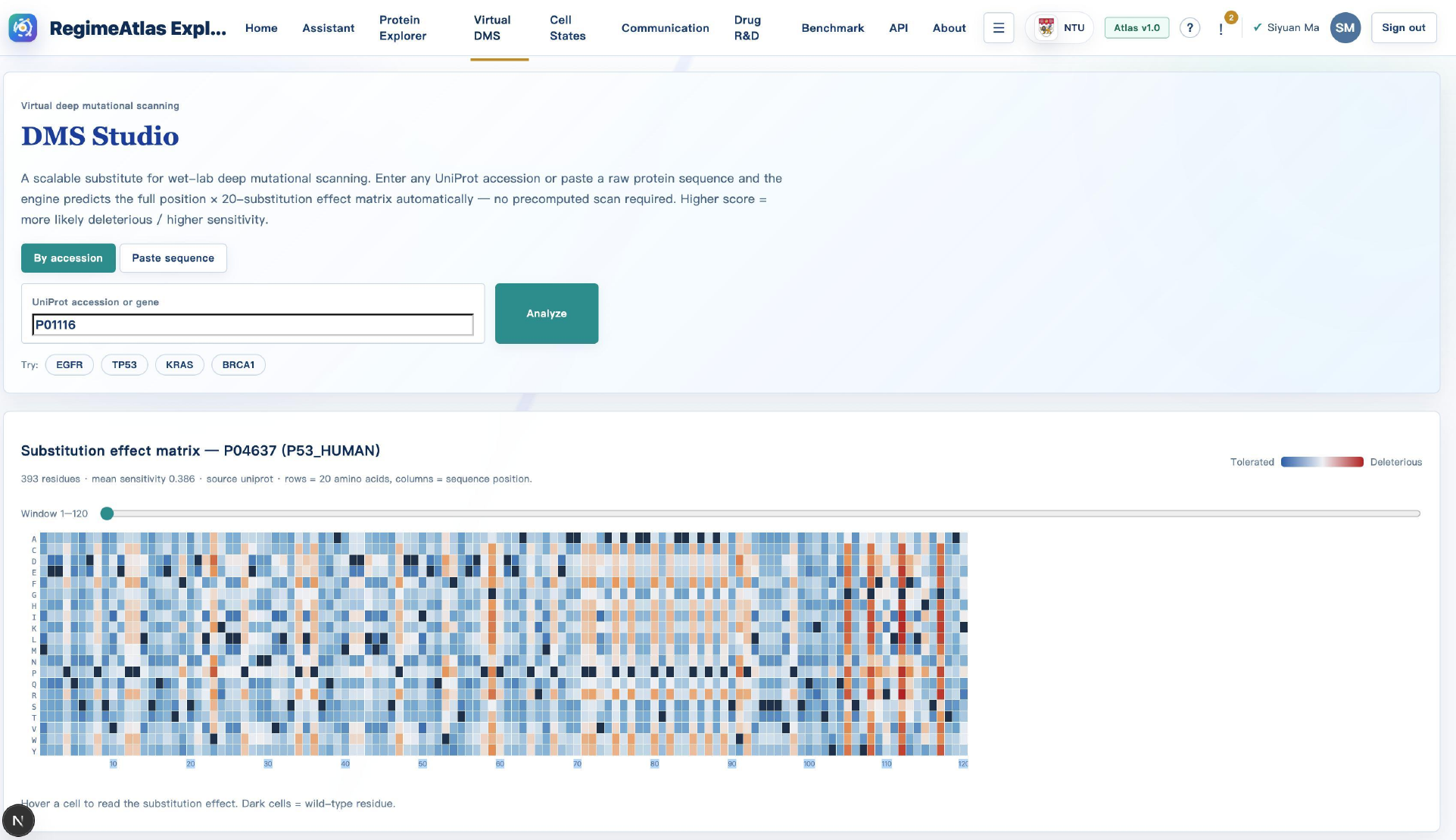}%
}
\end{center}
\end{landscape}
\clearpage
\captionsetup{font=small,justification=justified,singlelinecheck=false}
\captionof*{figure}{\textbf{Extended Data Fig. 8 | Interactive Virtual DMS Studio.} The DMS Studio accepts a UniProt accession or raw protein sequence and renders the substitution-effect matrix on demand. For TP53 (P04637), the displayed example covers 393 residues, reports mean sensitivity 0.386 and visualizes the 20-amino-acid substitution space across the selected sequence window. Together with Extended Data Fig. 6 and Supplementary Figs. 11--12, the matrix view shows the same perturbation object at mutation, residue and structural resolutions.}
\clearpage
\subsection*{Extended Data Table 1: perturbation regimes replicate across protein-model backbones}
This analysis tests whether the regime coordinate is tied to a particular ESM representation. Pairwise regime-score agreement is 0.742 for ESM2 versus ESM-1v, 0.688 for ESM2 versus ProtT5 and 0.671 for ESM-1v versus ProtT5, with a mean Spearman correlation of 0.700. Bin-level agreement remains moderate-to-strong ($\kappa=0.57$--0.64), and the direction of biological effects agrees in 91--94\% of comparisons, whereas shuffled or length-only controls remain at 0.06--0.08. The combination of continuous-coordinate, bin-level and effect-direction agreement argues that the learned perturbational organization is shared across representation families rather than inherited from a single encoder geometry.

\begin{table}[H]
\centering
\small
\caption*{\textbf{Extended Data Table 1 | Cross-backbone regime replication.}}
\begin{tabularx}{\textwidth}{Xcccc}
\toprule
Backbone pair & Regime $\rho$ & Cohen's $\kappa$ & Effect-direction agreement & Control agreement \\
\midrule
ESM2 vs. ESM-1v & 0.742 & 0.64 & 94\% & 0.08 \\
ESM2 vs. ProtT5 & 0.688 & 0.59 & 92\% & 0.07 \\
ESM-1v vs. ProtT5 & 0.671 & 0.57 & 91\% & 0.06 \\
Mean & 0.700 & 0.60 & 92\% & 0.07 \\
\bottomrule
\end{tabularx}
\par\vspace{0.45em}\footnotesize Regime-score Spearman $\rho$ quantifies agreement of continuous regime coordinates across representation backbones; Cohen's $\kappa$ measures agreement after discretization into regime bins. Effect-direction agreement summarizes concordance of the associated biological effects, and control agreement is calculated from shuffled and length-only representations.
\end{table}

\clearpage
\subsection*{Extended Data Table 2: regime-aware prediction transfers across laboratories and assay protocols}
Cross-laboratory transfer asks whether assay-conditioned prediction survives a protocol change rather than only a protein hold-out. Training or calibrating on Lab A and testing on Lab B gives Spearman $\rho=0.612$ for RegimeFormer versus 0.503 for PLM-only prediction; the reverse direction gives 0.598 versus 0.481. Calibration slopes remain close to unity for RegimeFormer (0.98 and 1.02) but shift farther for the PLM-only model (0.87 and 0.84). The regime-aware model therefore retains a median 85\% of within-laboratory rank performance, compared with about 70\% for PLM-only prediction. This transfer result complements the unseen-family and temporal tests by showing robustness to experimental protocol as a distinct source of distribution shift.

\begin{table}[H]
\centering
\small
\caption*{\textbf{Extended Data Table 2 | Cross-laboratory and cross-protocol transfer.}}
\begin{tabularx}{\textwidth}{p{0.17\textwidth}Xccc}
\toprule
Direction & Model & Cross-lab $\rho$ & Calibration slope & $R_{\mathrm{transfer}}$ \\
\midrule
A $\rightarrow$ B & RegimeFormer & 0.612 & 0.98 & 0.86 \\
A $\rightarrow$ B & PLM-only & 0.503 & 0.87 & 0.71 \\
B $\rightarrow$ A & RegimeFormer & 0.598 & 1.02 & 0.83 \\
B $\rightarrow$ A & PLM-only & 0.481 & 0.84 & 0.69 \\
Median (RegimeFormer) & -- & 0.605 & 1.00 & 0.85 \\
\bottomrule
\end{tabularx}
\par\vspace{0.45em}\footnotesize $R_{\mathrm{transfer}}=\rho_{\mathrm{cross-lab}}/\rho_{\mathrm{within-lab}}$. Reciprocal A$\rightarrow$B and B$\rightarrow$A transfers were evaluated with the same held-out protocol and calibration procedure.
\end{table}

\end{document}